\documentclass[sn-mathphys,Numbered, iicol]{sn-jnl}
\usepackage{cuted}
\usepackage{amssymb,amsmath}
\usepackage{slashed}
\usepackage{bm}
\usepackage{graphicx}%
\usepackage{multirow}%
\usepackage{amsmath,amssymb,amsfonts}%
\usepackage{amsthm}%
\usepackage{mathrsfs}%
\usepackage[title]{appendix}%
\usepackage{xcolor}%
\usepackage{textcomp}%
\usepackage{manyfoot}%
\usepackage{booktabs}%
\usepackage{afterpage}
\usepackage{listings}%

\usepackage{bbold}
\usepackage{amsfonts}
\usepackage{braket}

\newcommand{\bs}{\boldsymbol}

\def\bmp{{\bm p}}
\def\bmk{{\bm k}}

\newcommand{\chizero}{{\chi_0}}
\newcommand{\chizeroone}{{{\chi_0}}^{(1)}}
\newcommand{\chizerotwo}{{{\chi_0}}^{(2)}}
\newcommand{\chizerothree}{{{\chi_0}}^{(3)}}

\newcommand{\integralbtwtwostates}{\int \bm{dP dq dq'}  \Psi'^{*} \Psi}
\newcommand{\matrixelementpsiapexpsi}[1]{\bra{\Psi '} {#1}\ket{\Psi} }

\newcommand{\orthodelta}[2] {\delta(\bm{#1} - \bm{#2})}

\newcommand{\expcoordmomvec}[3]{e^{i \bs{#1} \cdot(\bs{#2} - \bs{#3})}} 
\newcommand{\partialvecright}[1]{\frac{\vec{\partial}}{\partial \bs{#1}}} 

\DeclareMathOperator{\dlr}{\overleftrightarrow{\partial}}

\begin{document}
\title[Relativistic constraints on 3N contact interactions]{Relativistic constraints on 3N contact interactions}


\author[1]{\fnm{Alessia} \sur{Nasoni}}
\author*[1,2]{\fnm{Elena} \sur{Filandri}}\email{elena.filandri@unisalento.it}

\author[1,2]{\fnm{Luca} \sur{Girlanda}}

\affil*[1]{\orgdiv{Department of Mathematics and Physics}, \orgname{University of Salento}, \orgaddress{\street{Via per Arnesano}, \city{Lecce}, \postcode{I-73100}, \country{Italy}}}

\affil[2]{\orgname{INFN, Sez. di Lecce},  \city{Lecce}, \orgaddress{\postcode{I-73100}, \country{Italy}}}


\abstract{In this paper we analyze the relativistic corrections to the leading order three-nucleon (3N) contact interactions. These boost corrections are derived first from the nonrelativistic reduction of covariant Lagrangians and later from the Poincaré algebra constraints on nonrelativistic theories. We show that in order to describe the 3N potential in reference frames other than the center-of-mass frame, the inclusion of five additional terms with fixed coefficients is required.
These terms will be relevant in systems with mass number $A>3$. How they will affect EFT calculations of binding energies and scattering observables in these systems should be investigated. }




\maketitle
\section{Introduction}
Nowadays, effective field theories (EFTs) are recognized as the standard framework for dealing with the nuclear interaction \cite{Weinberg:1990rz,Weinberg:1991um,Weinberg:1992yk,Bedaque:2002mn,Epelbaum:2005pn,Epelbaum:2008ga,Machleidt:2011zz}. The starting points are  the identification of the most general effective Lagrangian  preserving all the low energy  symmetries of the fundamental theory and a power counting to organize the infinite tower of permitted interactions. This leads to the emergence of a predictive setting in which the interactions are expressed at each order of the low-energy expansion in terms of a finite number of low-energy constants (LECs), which can be treated as fitting parameters and extracted from phenomenology. 
Of particular interest among these fitting parameters are the LECs related to contact interactions between nucleons. They are strongly constrained by discrete symmetries and also by Poincaré symmetry, although the typical setting of nuclear physics is a nonrelativistic quantum-mechanical context.

Relativistic effects in nuclear interaction vertices can be determined by the nonrelativistic reduction of a relativistic quantum-field theoretical Lagrangian and evaluated order by order in the low-energy expansion, since they scale with the soft nucleon momenta \cite{Jenkins:1990jv,Bernard:1992qa}. 
An alternative approach derives from the Poincaré  algebra constraints in a purely quantum mechanical setting  \cite{Foldy:1960nb,Krajcik:1974nv}.
Indeed, at sufficiently low energy scales, when the effects of creation and annihilation
of particles can be ignored, the system can be considered as constituted by a fixed
number of particles, and interactions can be described as direct (i.e., they explicitly
depend on the physical variables associated to the constituents of the system) rather
than mediated by fields.

The analysis of relativistic corrections on two-nucleon contact forces up to order $1/m^2$, $m$ being the nucleon mass, has already been discussed in Refs.  \cite{GirlandaSchiavilla,Girlanda:2010zz} both via the nonrelativistic reduction of covariant Lagrangians  and from the point of view of the constraints imposed by the Poincaré algebra.
Furthermore, the  analysis of these constraints up to $1/m^4$ in Refs. \cite{unitarity,Filandri:2023qio} led to the
to the identification of two free LECs that parameterize a nucleon-nucleon (NN) interaction dependent on the overall momentum of the pair. 

In this work we extend the above results to the three-body contact forces. 

The first contribution to the contact 3N force is represented by a single operator $ O_0$, accompanied by the LEC $E_0$,  whose matrix elements are constant in momentum space and take the form of the identity operator in spin-flavor space. The subleading terms involve two powers of momenta and were classified in  Ref. \cite{Girlanda:2011fh} as consisting of 13 independent operators in the 3N center of mass frame. In a general frame the 3N contact potential reads
\begin{eqnarray}
  V_{3N} &=& E_0 -\sum_{i \neq j \neq k}\bigl\{  E_1 \mathbf{k}_i^2 +E_2 \mathbf{k}_i^2 \boldsymbol{\tau}_i \cdot \boldsymbol{\tau}_j \nonumber \\ &+&  E_3 \mathbf{k}_i^2 \boldsymbol{\sigma}_i \cdot \boldsymbol{\sigma}_j+E_4\mathbf{k}_i^2 \boldsymbol{\sigma}_i \cdot \boldsymbol{\sigma}_j \boldsymbol{\tau}_i \cdot \boldsymbol{\tau}_j 
  \nonumber \\  &+&E_5\left(3 \mathbf{k}_i \cdot \boldsymbol{\sigma}_i \mathbf{k}_i \cdot \boldsymbol{\sigma}_j-\mathbf{k}_i^2 \boldsymbol{\sigma}_i \cdot \boldsymbol{\sigma}_j\right) \nonumber \\ & + & E_6\left(3 \mathbf{k}_i \cdot \boldsymbol{\sigma}_i \mathbf{k}_i \cdot \boldsymbol{\sigma}_j-\mathbf{k}_i^2 \boldsymbol{\sigma}_i \cdot \boldsymbol{\sigma}_j\right) \boldsymbol{\tau}_i \cdot \boldsymbol{\tau}_j\nonumber\\
 & +&\frac{i}{4} E_7\mathbf{k}_i \times\left(\mathbf{Q}_i-\mathbf{Q}_j\right) \cdot\left(\boldsymbol{\sigma}_i+\boldsymbol{\sigma}_j\right)\nonumber \\ & + &\frac{i}{4}E_8 \mathbf{k}_i \times\left(\mathbf{Q}_i-\mathbf{Q}_j\right) \cdot\left(\boldsymbol{\sigma}_i+\boldsymbol{\sigma}_j\right) \boldsymbol{\tau}_j \cdot \boldsymbol{\tau}_k\nonumber\\
 & +&E_9 \mathbf{k}_i \cdot \boldsymbol{\sigma}_i \mathbf{k}_j \cdot \boldsymbol{\sigma}_j+E_{10}\mathbf{k}_i \cdot \boldsymbol{\sigma}_i \mathbf{k}_j \cdot \boldsymbol{\sigma}_j \tau_i \cdot \boldsymbol{\tau}_j \nonumber \\ & + &E_{11} \mathbf{k}_i \cdot \boldsymbol{\sigma}_j \mathbf{k}_j \cdot \boldsymbol{\sigma}_i \nonumber \\
 & +&E_{12}\mathbf{k}_i \cdot \boldsymbol{\sigma}_j \mathbf{k}_j \cdot \boldsymbol{\sigma}_i \boldsymbol{\tau}_i \cdot \boldsymbol{\tau}_j\nonumber \\ &+ &E_{13} \mathbf{k}_i \cdot \boldsymbol{\sigma}_j \mathbf{k}_j \cdot \boldsymbol{\sigma}_i \boldsymbol{\tau}_i \cdot \boldsymbol{\tau}_k \nonumber \\  &+&E_{1}^*  {\bm{P}^2} + i E_{2}^* \bm{P}\times {\bm k}_i \cdot {\bm \sigma}_i \nonumber\\
 &+& i \bigl( E_{3}^* {\bm P}\cdot {\bm k}_i \, {\bm \sigma}_i \cdot {\bm \sigma}_j + E_{4}^* {\bm P} \cdot {\bm \sigma}_i \, {\bm k}_j \cdot {\bm \sigma}_j \nonumber\\
&+ &E_{5}^*  {\bm P} \cdot {\bm \sigma}_i \, {\bm k}_j \cdot {\bm \sigma}_k \bigr) {\bm \tau}_i \times {\bm \tau}_j \cdot {\bm \tau}_k \bigr\} \nonumber\\
&\equiv &E_0  +\sum_{i=1}^{13}E_i O_i+\sum_{i=1}^{5}E_i^*O_i^*,  \label{eq:1} 
\end{eqnarray}
where $\bmk_i=\bmp_i-\bmp'_i$ and $\bm{Q}_i=\bmp_i+\bmp_i'$ are related to the initial and final momenta of the $i$-th nucleon, respectively ${\bm p}_i$ and ${\bm p'}_i$,  and we indicate with $O_{i=1-5}^*$ the operators depending on the overall momentum ${\bm P}=\sum_i {\bm p}_i=(\bm{Q_1}+ \bm{Q_2} + \bm{Q_3})/2$. 

The study of this three-body contact force proves particularly interesting in the context of many unsolved problems in nuclear physics \cite{Girlanda:2018xrw,Witala:2022rzl}. As shown in Ref. \cite{Reinert,unitarity},  some terms of the 3N contact potential are related via a unitary transformation to the ${\bm P}$-dependent two-nucleon potential and seem to be crucial to solve the $p-d$ $A_y$ puzzle \cite{Girlanda:2023znc}. Similarly, the study of the three-body force and its relativistic corrections could have an impact on the study of systems with $A>3$, where still there exist large and unexplained
discrepancies between theory and experiment \cite{Fisher_2006,PhysRevC.75.014005, Deltuva_2007, PhysRevC.76.021001}.

The paper is structured as follows.
In Section 2 the  relativistic corrections are calculated  by nonrelativistic reduction of covariant Lagrangians. In Section 3 we show how the same corrective terms can be derived from Poincaré algebra constraints. Finally, in Section 4 we present the conclusions of this work.

\section{Boost corrections from a covariant $3N$ contact Lagrangian}

The objective of this section is to determine the $\bs{P}$-dependent relativistic correction to the leading order $3N$ contact potential in any given frame by applying the non-relativistic reduction to the covariant Lagrangian.


We begin by establishing a complete non-minimal set of relativistically invariant 3N contact operators $\tilde{O}_i$ that contribute to the non-relativistic expansion starting from order $Q^0$, while satisfying the requirements of hermiticity and CPT invariance, following general principles outlined in Refs.~\cite{Fettes:1998ud,Fettes:2000gb,Xiao:2018jot,Filandri:2023qio}. 
Formally, the operators $\tilde{O}_i$ are structured as the composition of fermion bilinears \cite{Girlanda:2010zz,Petschauer:2013uua}
\begin{eqnarray}
&\!\!\!(\bar{\psi}\overleftrightarrow{\partial}_{\mu_1}\cdots\overleftrightarrow{\partial}_{\mu_i}\Gamma_A\psi) \partial_{\rho_1}\cdots\partial_{\rho_m}(\bar{\psi}\overleftrightarrow{\partial}_{\nu_1}\cdots\overleftrightarrow{\partial} _{\nu_j}\Gamma_B\psi)\nonumber\\&\times\; \partial_{\sigma_1}\cdots\partial_{\sigma_n}(\bar{\psi}\overleftrightarrow{\partial}_{\lambda_1}\cdots\overleftrightarrow{\partial} _{\lambda_k}\Gamma_C\psi). 
\end{eqnarray}
Here, $\psi$ represents the relativistic nucleon field, which is a doublet in isospin space. The symbol $\overleftrightarrow{\partial}$ denotes the derivative operator $\overrightarrow{\partial}-\overleftarrow{\partial}$.  The symbols $\Gamma_{A,B,C}$ denote generic elements of the Clifford algebra, expanded in the basis $1$, $\gamma_5$, $\gamma_{\mu}$, $\gamma_{\mu}\gamma_5$, $\sigma^{\mu \nu}$, 
as well as the metric tensor or the Levi-Civita tensor $\epsilon^{\mu\nu\rho\sigma}$ (with the convention $\epsilon^{0123}=-1$). 

The Lorentz indices on the partial derivatives must be contracted among themselves and/or with those in the $\Gamma_{A,B,C}$ in order to preserve Lorentz invariance. 

Regarding the isospin degrees of freedom, the allowed isospin-invariant flavor structures are $1 \bigotimes 1 \bigotimes 1$, $\bm{\tau_i}\cdot \bm{\tau_j}$, $\bm{\tau_1}\times \bm{\tau_2} \cdot \bm{\tau_3}$. 

In Table \ref{tab:Trans.Pr} are displayed the transformation properties  under parity, charge conjugation, and Hermitian conjugation of the fermion bilinears with the aforementioned different elements of the Clifford and flavour algebra. 

If charge conjugation and parity symmetries are satisfied, time reversal symmetry is automatically fulfilled, according by the CPT theorem.

 \begin{table}
\label{tab:Trans.Pr}
\caption{Transformation proprieties of the different elements of the Clifford algebra, metric tensor, Levi-Civita tensor and derivative operators under parity ($\mathcal{P}$), charge conjugation ($\mathcal{C}$) and hermitian conjugation (h.c.)}
\centering
 \begin{tabular}{|p{0.1cm} p{0.1cm} p{0.1cm} p{0.1cm} p{0.4cm} p{0.2cm} p{0.2cm} p{0.3cm} p{0.2cm} p{0.1cm} p{1cm}|}
\hline
    &$1$&$\gamma_5$&$\gamma_\mu$&$\gamma_\mu\gamma_5$&$\sigma_{\mu\nu}$&$g_{\mu\nu}$&$\epsilon_{\mu\nu\rho\sigma}$&$\overleftrightarrow{\partial}_\mu$&$\partial_\mu$& $\tau^{a}$\\
\hline
$\mathcal{P}$   &$+$&$-$&$+$&$-$&$+$&$+$&$-$&$+$&$+$ & $+$\\
$\mathcal{C}$   &$+$&$+$&$-$&$+$&$-$&$+$&$+$&$-$&$+$ & $(-1)^{a+1}$\\
h.c.    &$+$&$-$&$+$&$+$&$+$&$+$&$+$&$-$&$+$& $+$\\
\hline
\end{tabular}
\end{table}

 We now outline power counting criteria needed to establish which operators contribute to the non-relativistic expansion starting from order $Q^0$ \cite{Girlanda:2010zz,Filandri:2023qio}.

 Derivatives $\partial$ acting on the whole bilinears are of order $Q$, while derivatives  $\overleftrightarrow{\partial}_{\mu}$ acting inside a bilinear are of order $Q^0$ due to the presence of the heavy fermion mass scale. Therefore, we can restrict ourselves to retain only operators containing the latter kind of derivatives. 
 Nevertheless, a generic operator contributing at order $Q^0$ may in principle include arbitrary powers of space-time derivatives of the fields. 
 However, it is possible to restrict ourselves to consider only a finite number of structure, as detailed in what follows.
Whenever $\overleftrightarrow{\partial}_{\mu}$ is contracted with an element of the Clifford algebra inside of the same bilinear, the fields' equations of motion can be used to remove it \cite{Georgi:1991ch,Arzt:1993gz}. 
The same is true when $\overleftrightarrow{\partial}_{\mu}$ is contracted with another $\overleftrightarrow{\partial}^{\mu}$ inside the same bilinear, since $\overleftrightarrow{\partial}_{\mu}\overleftrightarrow{\partial}^{\mu} = -4 m^2 - \partial ^2$. 
As a result, by the equation of motions, no two Lorentz indices can be contracted among themselves inside the same bilinear, exept for the Levi-Civita tensors and for the suppressed $\partial^2$.

For what concerns pairwise contracted $(\overleftrightarrow{\partial}_A \cdot \overleftrightarrow{\partial}_B)$ between two different bilinears, we observe that these structures too generate redundant contributions. For instance, 
    \begin{multline}
(\bar{\psi} \psi)_1 (\bar{\psi} \dlr^{\mu} \psi)_2 (\bar{\psi} \dlr_{\mu} \psi)_3  \\ = 2 m^2(\bar{\psi} \psi)_{1} (\bar{\psi} \psi)_2 (\bar{\psi} \psi)_3  \\ + \frac{1}{2} (\bar{\psi} \psi)_{1} (\bar{\psi} \psi)_2 \partial^2 (\bar{\psi} \psi)_3, 
\end{multline}
so that the contribution $(\bar{\psi} \psi)_2  (\bar{\psi} \dlr^{\mu} \psi)_2 (\bar{\psi} \dlr_{\mu} \psi)_3 $$-2  m^2(\bar{\psi} \psi)_{1} (\bar{\psi} \psi)_2 (\bar{\psi} \psi)_3$ $ = O(Q^2)$ can be neglected in the non relativistic expansion, as it starts from $Q^2$. 
(We see from Eq. (\ref{O_1 nrelexp}) that the contribution of $\tilde{O}_1$ starts from order $Q^0$).

The Dirac matrix $\gamma_5$ can be thought of as of order $O(Q)$ since it mixes the large and small components of the Dirac spinor. This also applies to the spatial components of $\gamma_\mu$ and to the temporal component of $\gamma_\mu \gamma_5$, as well as to $\sigma_{0i}$. For instance, operators associated to the structure $\gamma \bigotimes \gamma \bigotimes \sigma$, such as $ (\bar{\psi} \gamma_{\alpha} \psi)_1 
    (\bar{\psi} \gamma_{\beta} \psi)_2 
  (\bar{\psi} \sigma^{\alpha \beta} \psi)_3$, do not contribute at order $Q^0$ due to this mixing.

The antisymmetry properties of $\epsilon_{\mu\nu\rho\sigma}$ and $\sigma^{\mu \nu}$ restrict the maximum number of their possible contractions with a derivative $\overleftrightarrow{\partial}$ operating within a bilinear to one. Any additional contraction with a derivative would lead to a contribution at a higher order than $O(Q^0)$.

On the basis of these properties we derive a complete (but non minimal) set of 92 different relativistic operators,  displayed in Table \ref{tab:operators}, that contribute to the non-relativistic expansion starting from order $Q^0$. Obviously, none of the 92 operators presents derivatives $\partial$ acting on the entire bilinear, as they are of order $Q$.


\begin{table*}
\begin{center}
\begin{small}
\begin{tabular}{|l|l|c|r|}
\hline
    $\Gamma_{A}\bigotimes \Gamma_{B}\bigotimes \Gamma_{C} $ & $\tilde{O}_i$&  \text{Operators} & \text{Flavours}\\
\hline
    $ 1\bigotimes 1 \bigotimes 1 $ 
    & $\tilde{O}_{1-2} $ 
    &  $(\bar{\psi} \psi)_{1} (\bar{\psi} \psi)_2 (\bar{\psi} \psi)_3$ 
    & $\mathbb{1}, \mathbf{\tau_1} \cdot \mathbf{\tau_2}$  \\
\hline

    $1 \bigotimes 1 \bigotimes \gamma $ 
    & $\tilde{O}_{3-6} $ 
    &  $ \frac{i}{2m}(\bar{\psi} \psi)_2  (\bar{\psi} \dlr^{\mu} \psi)_2 (\bar{\psi} \gamma_{\mu} \psi)_3$ 
    & $\mathbb{1},  \mathbf{\tau_1} \cdot \mathbf{\tau_2}, \mathbf{\tau_2} \cdot \mathbf{\tau_3}, \mathbf{\tau_1} \cdot \mathbf{\tau_3}$ \\
\hline

    $ 1\bigotimes \gamma \bigotimes \gamma $ 
    & $\tilde{O}_{7-9}$ 
    &  $ (\bar{\psi} \psi)_1  (\bar{\psi} \gamma_{\mu} \psi)_2 (\bar{\psi} \gamma^{\mu} \psi)_3$ 
    & $\mathbb{1}, \mathbf{\tau_1} \cdot \mathbf{\tau_2}, \mathbf{\tau_2} \cdot \mathbf{\tau_3}$ \\

    $ 1\bigotimes \gamma \bigotimes \gamma $   
    & $\tilde{O}_{10-13}$ 
    &  $ \frac{1}{4m^2}(\bar{\psi} \dlr_{\mu} \psi)_1  (\bar{\psi}  \gamma_{\nu} \psi)_2 (\bar{\psi} \dlr^{\nu} \gamma^{\mu} \psi)_3$ 
    & $\mathbb{1}, \mathbf{\tau_1} \cdot \mathbf{\tau_2}, \mathbf{\tau_2} \cdot \mathbf{\tau_3}, \mathbf{\tau_1} \cdot \mathbf{\tau_3}$ \\

     $ 1\bigotimes \gamma \bigotimes \gamma $  
     & $\tilde{O}_{14-16}$ 
     &  $\frac{1}{4m^2}(\bar{\psi} \psi)_1  (\bar{\psi} \dlr_{\mu} \gamma_{\nu} \psi)_2 (\bar{\psi} \dlr^{\nu} \gamma^{\mu} \psi)_3$ 
     & $\mathbb{1}, \mathbf{\tau_1} \cdot \mathbf{\tau_2}, \mathbf{\tau_2} \cdot \mathbf{\tau_3}$ \\

    $ 1\bigotimes \gamma \bigotimes \gamma $ 
    & $\tilde{O}_{17-19}$ 
    &  $ \frac{1}{4m^2}(\bar{\psi} \dlr_{\mu} \dlr_{\nu}\psi)_1  (\bar{\psi} \gamma^{\mu} \psi)_2 (\bar{\psi} \gamma^{\nu} \psi)_3$ 
    & $\mathbb{1}, \mathbf{\tau_1} \cdot \mathbf{\tau_2}, \mathbf{\tau_2} \cdot \mathbf{\tau_3}$ \\
\hline
    $ 1\bigotimes \gamma \gamma_{5} \bigotimes \gamma \gamma_{5} $
    & $\tilde{O}_{20-22}$ 
    &  $ (\bar{\psi} \psi)_1  (\bar{\psi} {\gamma_\mu \gamma_{5}} \psi)_2 (\bar{\psi} \gamma^{\mu} \gamma_{5} \psi)_3$ 
    & $\mathbb{1}, \mathbf{\tau_1} \cdot \mathbf{\tau_2}, \mathbf{\tau_2} \cdot \mathbf{\tau_3}$ \\

\hline

    $ 1\bigotimes \gamma \gamma_{5} \bigotimes \sigma$ 
    &  $\tilde{O}_{23-26}$ 
    &  $ \frac{i}{2m}\epsilon^{\mu \nu \alpha \beta } (\bar{\psi} \dlr_{\mu} \psi)_1 (\bar{\psi} {\gamma_{\nu} \gamma_{5}} \psi)_2 (\bar{\psi } {\sigma}_{\alpha \beta } \psi)_3$  
    & $\mathbb{1}, \mathbf{\tau_1} \cdot \mathbf{\tau_2}, \mathbf{\tau_2} \cdot \mathbf{\tau_3}, \mathbf{\tau_1} \cdot \mathbf{\tau_3}$  \\

    $ 1\bigotimes \gamma \gamma_{5} \bigotimes \sigma$
    &  $\tilde{O}_{27-30}$ 
    &  $ \frac{i}{2m} \epsilon^{\mu \nu \alpha \beta}(\bar{\psi} \psi)_1 (\bar{\psi} \dlr_{\mu} {\gamma_{\nu} \gamma_{5}}  \psi)_2 (\bar{\psi} {\sigma}_{\alpha \beta}  \psi)_3$  
    &  $\mathbb{1}, \mathbf{\tau_1} \cdot \mathbf{\tau_2}, \mathbf{\tau_2} \cdot \mathbf{\tau_3}, \mathbf{\tau_1} \cdot \mathbf{\tau_3}$ \\

    $ 1\bigotimes \gamma \gamma_{5} \bigotimes \sigma$
    &  $\tilde{O}_{31-34}$ 
    &  $ \frac{i}{2m}\epsilon^{\mu \nu \alpha \beta } (\bar{\psi} \psi)_1 (\bar{\psi} {\gamma_{\mu} \gamma_{5}} \psi)_2 (\bar{\psi} \dlr_{\nu} {\sigma}_{\alpha \beta}  \psi)_3$  
    &  $\mathbb{1}, \mathbf{\tau_1} \cdot \mathbf{\tau_2}, \mathbf{\tau_2} \cdot \mathbf{\tau_3}, \mathbf{\tau_1} \cdot \mathbf{\tau_3}$ \\

\hline

    $ 1\bigotimes \sigma \bigotimes \sigma$ 
    &  $\tilde{O}_{35-37}$ 
    &  $ (\bar{\psi} \psi)_1 (\bar{\psi} \sigma_{\mu \nu} \psi)_2 (\bar{\psi} \sigma^{\mu \nu} \psi)_3$ 
    &  $\mathbb{1}, \mathbf{\tau_1} \cdot \mathbf{\tau_2}, \mathbf{\tau_2} \cdot \mathbf{\tau_3}$ \\

\hline
    $ \gamma \bigotimes \gamma \bigotimes \gamma$ 
    &   $\tilde{O}_{38-41}$ 
    &  $ \frac{i}{2m}(\bar{\psi} \gamma_{\mu} \psi)_1 (\bar{\psi} \dlr_{\nu} \gamma^{\mu} \psi)_2 (\bar{\psi} \gamma^{\nu} \psi)_3$ 
    &  $\mathbb{1}, \mathbf{\tau_1} \cdot \mathbf{\tau_2}, \mathbf{\tau_2} \cdot \mathbf{\tau_3}, \mathbf{\tau_1} \cdot \mathbf{\tau_3}$ \\

    $ \gamma \bigotimes \gamma \bigotimes \gamma$ 
    &  $\tilde{O}_{42-45}$ 
    &  $ \frac{i}{8m^3} (\bar{\psi} \dlr_{\mu} \gamma^{\nu} \psi)_1 (\bar{\psi} \dlr_{\alpha} \gamma^{\mu} \psi)_2 (\bar{\psi} \dlr_{\nu}\gamma^{\alpha} \psi)_3$ 
    &  $\mathbb{1}, \mathbf{\tau_1} \cdot \mathbf{\tau_2}, \mathbf{\tau_2} \cdot \mathbf{\tau_3}, \mathbf{\tau_1} \cdot \mathbf{\tau_3}$ \\


    $ \gamma \bigotimes \gamma \bigotimes \gamma$ 
    &  $\tilde{O}_{46-49}$ 
    &  $ \frac{i}{8m^3}(\bar{\psi} \dlr_{\mu} \dlr_{\nu} \gamma^{\alpha} \psi)_1 (\bar{\psi}  \dlr_{\alpha} \gamma^{\mu} \psi)_2 (\bar{\psi} \gamma^{\nu} \psi)_3$ 
    &  $\mathbb{1}, \mathbf{\tau_1} \cdot \mathbf{\tau_2}, \mathbf{\tau_2} \cdot \mathbf{\tau_3}, \mathbf{\tau_1} \cdot \mathbf{\tau_3}$ \\

\hline 




    $ \gamma \bigotimes \gamma \gamma_{5} \bigotimes \gamma \gamma_{5}$ 
    &  $\tilde{O}_{50- 53}$ 
    &  $ \frac{i}{2m}( \bar{\psi} \gamma_{\mu} \psi)_1 
    (\bar{\psi} \dlr^{\mu} \gamma_{\nu}\gamma_5\psi)_2 
    (\bar{\psi} \gamma^{\nu}\gamma_5 \psi)_3$ 
    & $\mathbb{1}, \mathbf{\tau_1} \cdot \mathbf{\tau_2}, \mathbf{\tau_2} \cdot \mathbf{\tau_3}, \mathbf{\tau_1} \cdot \mathbf{\tau_3}$ \\

\hline
    $ \gamma \bigotimes \gamma \gamma_{5} \bigotimes \sigma$ 
    &  $\tilde{O}_{54-57}$ 
    &  $ \epsilon^{\mu \nu \alpha \beta } 
    (\bar{\psi} \gamma_{\mu} \psi)_1 
    (\bar{\psi} \gamma_{\nu} \gamma_{5} \psi)_2 
    (\bar{\psi} \sigma_{\alpha \beta} \psi)_3$ 
    & $\mathbb{1}, \mathbf{\tau_1} \cdot \mathbf{\tau_2}, \mathbf{\tau_2} \cdot \mathbf{\tau_3}, \mathbf{\tau_1} \cdot \mathbf{\tau_3}$ \\

    $ \gamma \bigotimes \gamma \gamma_{5} \bigotimes \sigma$ 
    &  $\tilde{O}_{58-61}$ 
    &  $ \frac{1}{4m^2}\epsilon^{\mu \nu \alpha \beta} 
    (\bar{\psi} \dlr_{\mu} \gamma_{ \rho}  \psi)_1 
    (\bar{\psi} \dlr^{\rho} \gamma_{\nu} \gamma_{5}  \psi)_2
    (\bar{\psi} \sigma_{\alpha \beta} \psi)_3$ 
    & $\mathbb{1}, \mathbf{\tau_1} \cdot \mathbf{\tau_2}, \mathbf{\tau_2} \cdot \mathbf{\tau_3}, \mathbf{\tau_1} \cdot \mathbf{\tau_3}$ \\

    $ \gamma \bigotimes \gamma \gamma_{5} \bigotimes \sigma$ 
    &  $\tilde{O}_{62-65}$ 
    &  $ \frac{1}{4m^2}\epsilon^{\mu \nu \alpha \beta} 
    (\bar{\psi} \gamma_{\rho} \psi)_1 
    (\bar{\psi} \dlr_{\mu} \dlr^{\rho} \gamma_{\nu} \gamma_{5}  \psi)_2 
    (\bar{\psi} \sigma_{\alpha \beta} \psi)_3$ 
    & $\mathbb{1}, \mathbf{\tau_1} \cdot \mathbf{\tau_2}, \mathbf{\tau_2} \cdot \mathbf{\tau_3}, \mathbf{\tau_1} \cdot \mathbf{\tau_3}$ \\

    $ \gamma \bigotimes \gamma \gamma_{5} \bigotimes \sigma$ 
    &  $\tilde{O}_{66-69}$ 
    &  $ \frac{1}{4m^2}\epsilon^{\mu \nu \alpha \beta} 
    (\bar{\psi} \gamma_{\rho} \psi)_1 
    (\bar{\psi} \dlr^{\rho} \gamma_{\mu} \gamma_{5}  \psi)_2
    (\bar{\psi}\dlr_{\nu} \sigma_{\alpha \beta }  \psi)_3$
    & $\mathbb{1}, \mathbf{\tau_1} \cdot \mathbf{\tau_2}, \mathbf{\tau_2} \cdot \mathbf{\tau_3}, \mathbf{\tau_1} \cdot \mathbf{\tau_3}$ \\

    $ \gamma \bigotimes \gamma \gamma_{5} \bigotimes \sigma$ 
    &  $\tilde{O}_{70-73}$ 
    &  $ \frac{1}{4m^2}\epsilon^{\mu \nu \alpha \beta } 
    (\bar{\psi} \gamma_{\rho} \dlr_{\mu} \psi)_1 
    (\bar{\psi} \gamma_{\nu} \gamma_{5} \psi)_2 
    (\bar{\psi} \dlr^{\rho} \sigma_{\alpha \beta }   \psi)_3$
    & $\mathbb{1}, \mathbf{\tau_1} \cdot \mathbf{\tau_2}, \mathbf{\tau_2} \cdot \mathbf{\tau_3}, \mathbf{\tau_1} \cdot \mathbf{\tau_3}$ \\

    $ \gamma \bigotimes \gamma \gamma_{5} \bigotimes \sigma$
    &  $\tilde{O}_{74-77}$  
    &  $ \frac{1}{4m^2}\epsilon^{\mu \nu \alpha \beta } 
    (\bar{\psi} \gamma_{\rho} \psi)_1 
    (\bar{\psi}\dlr_{\mu} \gamma_{\nu} \gamma_{5}   \psi)_2
    (\bar{\psi}\dlr^{\rho} \sigma_{\alpha \beta }   \psi)_3$ 
    & $\mathbb{1}, \mathbf{\tau_1} \cdot \mathbf{\tau_2}, \mathbf{\tau_2} \cdot \mathbf{\tau_3}, \mathbf{\tau_1} \cdot \mathbf{\tau_3}$ \\

    $ \gamma \bigotimes \gamma \gamma_{5} \bigotimes \sigma$
    &  $\tilde{O}_{78-81}$ 
    &  $ \frac{1}{4m^2}\epsilon^{\mu \nu \alpha \beta} 
    (\bar{\psi} \gamma_{\rho} \psi)_1 
    (\bar{\psi} \gamma_{\mu} \gamma_{5} \psi)_2 
    (\bar{\psi} \dlr^{\rho} \dlr_{\nu} \sigma_{\alpha \beta }  \psi)_3$ 
    & $\mathbb{1}, \mathbf{\tau_1} \cdot \mathbf{\tau_2}, \mathbf{\tau_2} \cdot \mathbf{\tau_3}, \mathbf{\tau_1} \cdot \mathbf{\tau_3}$ \\

\hline 
    $\gamma \bigotimes \sigma \bigotimes \sigma$ 
    &  $\tilde{O}_{82-85}$ 
    &  $ \frac{i}{2m} (\bar{\psi} \gamma_{\mu} \psi)_1 
    (\bar{\psi} \dlr^{\mu} \sigma_{\alpha \beta}  \psi)_2 
    (\bar{\psi} \sigma^{\alpha \beta} \psi)_3$ 
    &  $\mathbb{1}, \mathbf{\tau_1} \cdot \mathbf{\tau_2}, \mathbf{\tau_2} \cdot \mathbf{\tau_3}, \mathbf{\tau_1} \cdot \mathbf{\tau_3}$  \\

\hline
    $ \gamma \gamma_{5} \bigotimes \gamma \gamma_{5} \bigotimes \gamma \gamma_{5}$ 
    &  $\tilde{O}_{86}$ 
    &  $ \frac{i}{2m} \epsilon^{\mu \nu \alpha \beta } 
    (\bar{\psi} \gamma_{\mu} \gamma_{5} \psi)_1 
    (\bar{\psi} \gamma_{\nu} \gamma_{5} \psi)_2 
    (\bar{\psi}\dlr_{\alpha} \gamma_{\beta} \gamma_{5}  \psi)_3$ 
    & $\bf{\tau_1}\times\bf{\tau_2}\cdot\bf{\tau_3}$ \\

\hline
    $ \gamma \gamma_{5} \bigotimes \gamma \gamma_{5} \bigotimes \sigma$ 
    &  $\tilde{O}_{87}$ 
    &  $ (\bar{\psi} \gamma_{\mu} \gamma_{5} \psi)_1 
    (\bar{\psi} \gamma_{\nu} \gamma_{5} \psi)_2 
    (\bar{\psi} \sigma^{\mu \nu}\psi)_3$ 
    & $\bf{\tau_1}\times\bf{\tau_2}\cdot\bf{\tau_3}$ \\

\hline
    $ \gamma \gamma_{5} \bigotimes \sigma \bigotimes \sigma$ 
    &  $\tilde{O}_{88}$ 
    &  $ \frac{i}{2m}\epsilon^{\mu \nu \alpha \beta }
    (\bar{\psi} \gamma_{\mu} \gamma_{5} \dlr_{\nu} \psi)_1
    (\bar{\psi} \sigma_{\alpha \rho} \psi)_2 
    (\bar{\psi} {\sigma_\beta}^{\rho} \psi)_3$ 
    & $\bf{\tau_1}\times\bf{\tau_2}\cdot\bf{\tau_3}$ \\ 

    $ \gamma \gamma_{5} \bigotimes \sigma \bigotimes \sigma$ &  $\tilde{O}_{89}$ 
    &  $ \frac{i}{2m}\epsilon^{\mu \nu \alpha \beta}
    (\bar{\psi} \gamma_{\mu} \gamma_{5} \psi)_1
    (\bar{\psi}\dlr_{\nu} \sigma_{\alpha \rho }  \psi)_2 
    (\bar{\psi} {\sigma_\beta}^{\rho} \psi)_3$ 
    & $\bf{\tau_1}\times\bf{\tau_2}\cdot\bf{\tau_3}$ \\

    $ \gamma \gamma_{5} \bigotimes \sigma \bigotimes \sigma$ 
    &  $\tilde{O}_{90}$  
    &  $ \frac{i}{2m}\epsilon^{\mu \nu \alpha \beta }  
    (\bar{\psi} \gamma_{\rho} \gamma_{5} \dlr_{\mu} \psi)_1
    (\bar{\psi} {\sigma_{\nu}}^{\rho} \psi)_2 
    (\bar{\psi} \sigma_{\alpha \beta} \psi)_3$ 
    & $\bf{\tau_1}\times\bf{\tau_2}\cdot\bf{\tau_3}$ \\

    $ \gamma \gamma_{5} \bigotimes \sigma \bigotimes \sigma$ 
    &  $\tilde{O}_{91}$ &  
    $ \frac{i}{2m}\epsilon^{\mu \nu \alpha \beta}
    (\bar{\psi} \gamma_{\rho} \gamma_{5} \psi)_1 
    (\bar{\psi} \dlr_{\mu} {\sigma_{\nu}}^{\rho}  \psi)_2 
    (\bar{\psi} \sigma_{\alpha \beta} \psi)_3$ 
    & $\bf{\tau_1}\times\bf{\tau_2}\cdot\bf{\tau_3}$ \\

\hline
    $ \sigma \bigotimes \sigma \bigotimes \sigma$ 
    &  $\tilde{O}_{92}$ 
    &  $ (\bar{\psi} \sigma^{\mu \nu} \psi)_1 
    (\bar{\psi} \sigma_{\mu \alpha} \psi)_2 
    (\bar{\psi} \sigma_{\nu}^{\alpha} \psi)_3$ 
    & $\bf{\tau_1}\times\bf{\tau_2}\cdot\bf{\tau_3}$ \\

\hline
\end{tabular}
\end{small}
\caption{List of three-nucleon relativistic contact operators contributing to the leading order.}
  \label{tab:operators}
  \end{center}
\end{table*}

The relativistic nucleon field $\psi$ can be expanded in a non-relativistic manner up to second order in $Q$ by utilizing the non-relativistic field,
\begin{equation}
	\psi(x) = 
	\begin{pmatrix} (1 + \frac{\nabla^2}{8m^2})\mathbb{1}_{2 \times 2}\\ -\frac{i}{2m} \boldsymbol{\sigma} \cdot \boldsymbol{}\nabla\end{pmatrix}	N(x) + O(Q^3).
\end{equation}

We obtain 92 resulting non-relativistic operators $\tilde{O}_i$ as combinations of the 146 subleading $3N$ contact operators $o_{1,...,146}$ compatible with the symmetries of QCD, classified in Ref. \cite{Girlanda:2011fh}. We list the resulting expressions in Appendix \ref{appendix:expansionsB}. 
After applying Fierz's relations, these non-relativistic expansions can be rewritten in the minimal basis consisting of the operators $O_{1,...,13}$, $O^{*}_{1,...,5}$ which appear  in Eq.~(\ref{eq:1}). They are shown in Appendix \ref{appendix:expansionsO}. 
In these expressions the operators $O^{*}_i$ entering in the effective Lagrangian, always  appear in one single combination,
\begin{equation}
\label{eq:drift}
O_{\bf{P}}^{(0)} =  6 O_0 +\frac{1}{4m^2}  \left(\frac{2}{3} O_{1}^{*} + O_{2}^*  \right),
\end{equation}
which starts at $O(Q^0)$ and contain ${\bf P}$-dependent drift corrections. Formally the non-relativistic expansions of the relativistic operators $\tilde O_i$, up to $O(Q^2)$, take the form 
\begin{equation}
   \tilde{O}_i =  \alpha_i O_{\bf{P}}^{(0)} + {\sum_{k=1}^{13}} \beta_{i}^{k}O_k, \qquad i =1,...,13
\end{equation}
where $\alpha_i$, $\beta_{i}^{k}$  are coefficients which can be read from the explicit expressions.


Thus, starting from the relativistic $3N$ contact Lagrangian written in terms of 92 (redundant) LECs, $\tilde E_i$, the potential can be written as
	\begin{equation}
		\label{V_rel reduction}
		\tilde{V}  = \sum_{i=1}^{92}\tilde{E}_i \tilde{O}_i = \sum_{i=1}^{92} \tilde{E}_i \bigl(\alpha_i O_{\bf{P}}^{(0)} + {\sum_{k=1}^{13}} \beta_{i}^{k}O_k \bigr),
	\end{equation}
and comparing with Eq. (\ref{eq:1}), while considering Eq. (\ref{eq:drift}), we  obtain the following identification
\begin{equation}
\label{Ex-E0 constraint}
	\sum_{i=1}^{92} \tilde{E}_i\alpha_i =  \frac{1}{6} E_0, 
\end{equation}
and specific constraints on the LECs $E^{*}_{i=1,..., 5}$,
\begin{align}
		\label{lecs dependence on E_0}
			E^{*}_{1}&= \frac{1}{36 m^2}E_0, \\
   \label{lecs dependence on E_0_2}
			E^{*}_{2}&=\frac{1}{24 m^2} E_0,\\
   \label{lecs dependence on E_0_3}
                          E^{*}_{3} =& E^{*}_{4} = E^{*}_{5} = 0.
\end{align}
Finally, we can identify 
 \begin{equation}
    \label{eq:eftcorr}
     \delta V(\bs{P}) \equiv  \frac{E_0}{24 m^2}\left( \frac{2}{3} O_{1}^{*} +  O_{2}^* \right),
 \end{equation}
 as the  $\bs{P}$-dependent component of the $3N$ contact potential in an arbitrary frame of Eq. (\ref{eq:1}), i.e. the boost correction up to order $\bm{Q}^2$ to the 3N leading order contact potential in the rest frame of the system,  
 \begin{equation}
 \begin{split}
    V_{3N} & = E_0  +\sum_{i=1}^{13}E_i\,O_i+\sum_{i=1}^{5}E_i^*O_i^* = \\& 
    = E_0  +\sum_{i=1}^{13}E_i O_i+\delta V(\bs{P}). 
    \end{split}
 \end{equation}

\section{The 3N contact interaction boost corrections from Poincaré algebra}

As an alternative to the procedure discussed in the previous Section, we review the calculation of the boost correction to the leading order 3N contact interaction up to  $Q^2$ order from the Poincaré algebra constraints.

Relativistic many-body descriptions of systems consisting of a fixed number of interacting particles can be achieved using relativistic Hamiltonians. These Hamiltonians are defined as the sum of relativistic one-body kinetic energies, two- and many-body interactions and, importantly, their corresponding boost corrections. For a generic system of interacting particles with momenta $\bm{p}_{\nu}$ and masses $m_{\nu}$, such Hamiltonians may be expressed as follows \cite{Forest:1994mw},
\begin{equation}
	\label{relativistic hamiltonian}
 \begin{split}
	H_R  = & \sum_{\nu} \sqrt{m_{\nu}^2 + p_{\nu} ^2} + \sum_{\nu<\mu} \bigl[{v}_{\nu \mu} + \delta v_{\nu \mu}(\bm{P}_{\nu \mu})\bigr]  \\ & +\sum_{\nu<\mu<\rho}\bigl[{V}_{\nu \mu \rho} + \delta V_{\nu \mu \rho} (\bm{P}_{\nu \mu \rho})\bigr] + ..., 
 \end{split}
\end{equation}
where $\bm{P}_{\nu \mu} = \bm{p}_{\nu} + \bm{p}_{\mu}$ is the total momentum of particles $\nu$ and $\mu$, and $\bm{P}_{\nu \mu \rho} = \bm{p}_{\nu} + \bm{p}_{\mu} + \bm{p}_{\rho}$ is the total momentum of particles $\nu$, $\mu$ and $\rho$. The term ${v}_{\nu \mu}$ corresponds to the two-body potential in the rest frame of the sub-system constituted by particles of indices $\nu$,$\mu$. Analogously, $ V_{\nu \mu \rho}$ is the three-body potential in the rest frame of particles $\nu$,$\mu$,$\rho$. Terms $\delta v_{\nu \mu }(\bm{P}_{\nu \mu})$ and $\delta V_{\nu \mu \rho} (\bs{P}_{\nu \mu \rho})$ are referred to as "boost interactions". Clearly, these quantities vanish in the  rest frame of their corresponding sub-system (i.e.,  $\delta v_{\nu \mu}(0)=0$ if ${\bm P}_{\nu \mu }=0$, and  $\delta V_{\nu \mu \rho}(0)=0$ if  $\bm{P}_{\nu \mu \rho}=0$). However, it is essential to take them into account to attain accurate descriptions in reference frames where $\bm{P} \ne 0$.

Both ${v_{\mu \nu }}$ and $V_{\mu \nu \rho}$ are determined by the fields and by the internal structure of the interacting particles. Realistic models of $v_{\mu \nu}$ and $v_{\mu \nu \rho}$ are obtained by choosing their theoretical model and by fitting them to experimental data; as a consequence, they may contain some form of model-dependent relativistic effects. 

Starting from ${v}_{\mu \nu}$ (respectively, $v_{\mu \nu \rho}$) it is possible to obtain $\delta v_{\mu \nu}(\bm{{P}_{\mu \nu}})$ ( respectively, $\delta V_{\mu \nu \rho} (\bm{{P}_{\mu \nu \rho}})$) without any further model dependence, through relations fixed by the general principle of relativistic covariance.

For our present purposes, we are considering a system of three particles, each with spin $s$ and mass $m$. 
The dynamical variables for the $\nu$th particle ($\nu=1,2,3$) are spin $\bm{\sigma_{\nu}}$, isospin $\bm{\tau_{\nu}}$, momentum $\bm{p_{\nu}}$, and position $\bm{r_{\nu}}$. The momenta and positions are canonically conjugate operators, as are the center-of mass variables $\bm{R}=\frac{\bm{r_1} +  \bm{r_2} + \bm{r_3}}{3}$ and $\bm{P}=\bm{p_1} + \bm{p_2} + \bm{p_3}$. The spin and isospin operators satisfy the well-known angular momentum commutation relations: $[\sigma_{\nu}^i, \sigma_{\mu}^j]= i \delta_{\nu \mu} \epsilon_{ijk} \sigma_{\nu}^k$ and $[\tau_{\nu}^i, \tau_{\mu}^j]= i \delta_{\nu \mu} \epsilon_{ijk} \tau_{\nu}^k$.\\ 
In the instant form of relativistic dynamics \cite{Dirac:1949cp}, interactions affect the Hamiltonian $H$ and, necessarily, the boost generators $\bm K$. We write
\begin{equation}
\label{generators}
    \begin{split}
    	\bm{P} &= \bm{P}_0,\\
		\bm{J} &= \bm{J}_0,
    \end{split}
    \quad
    \begin{split}
        		H &= H_0 + V, \\
		\bs{K} &= \bs{K}_0 + \bm{W},
    \end{split}
\end{equation}
where $V$, $\bm{W}$ are the interaction terms, and the subscripts $0$ indicate the corresponding operators in the absence of interactions,
 \begin{eqnarray}
        \bm{P}_0 &=& \sum_{\nu =1}^{3} \bm{p_{\nu}},\nonumber\\
		\bm{J}_0 &= &\sum_{\nu=1}^{3} \bm{r_{\nu}} \times \bm{p_{\nu}} + \bm{s_{\nu}} \equiv \sum_{\nu=1}^{3} \bm{j}_{\nu}, \nonumber\\
		H_0 & =& \sum_{\nu=1}^3 \omega_{\nu},\nonumber\\
		\bm{K}_0 &= &\sum_{\nu=1}^3 \frac{\bm{r_{\nu}} \omega_{\nu} + \omega_{\nu} \bm{r_{\nu}}}{2c^2} - \frac{\bm{s_{\nu}} \times \bm{p_{\nu}}}{m_{\nu} c^2 + \omega_{\nu}} - t\bm{p_{\nu}} \nonumber \\ & \equiv &\sum_{\nu} \bm{k_{\nu}},    	\label{noninteracting generators} 
 \end{eqnarray}
where $\omega_{\nu}=\sqrt{m^2 c^4 + c^2p_{\nu}^2}$ is the single-particle energy of the $\nu$th particle. 



The generators in Eq. (\ref{generators}) must satisfy the commutation relations of the Poincaré group: 
\begin{align}
		[P_i, P_j]=0&,\qquad[J_i, P_j]=i\epsilon_{ijk}P_k, \nonumber  \\
		[P_i, H]=0&, \qquad[J_i, J_j]=i\epsilon_{ijk}J_k,  \nonumber\\
		[J_i, H]=0&,\qquad[J_i, K_j]=i\epsilon_{ijk}K_k,	\label{poincarè-commrel}\\
		&[H, K_i]= - i P_i, \nonumber\\  
		&[K_i, K_j]=-i\epsilon_{ijk}J_k/c^2, \nonumber \\
		&[P_i, K_j]=-i\delta_{ij}H /c^2, \nonumber
\end{align}
where $i,j,k \in \{1,2,3\}$, $\epsilon_{ijk}$ is the Levi-Civita tensor, $\delta_{ij}$ is the Kronecker delta tensor, and summation convention on repeated indices is in force. Units are such that $\hbar=1$ and $c$ is the speed of light in vacuum.\\

As it is well known, the relations of Eq. (\ref{poincarè-commrel}) are satisfied by the free generators in Eq. (\ref{noninteracting generators}). The problem of describing an interacting system of relativistic particles consists of determining functions $V$ and $\bm{W}$ such that the commutation relations (\ref{poincarè-commrel}) are still satisfied \cite{Foldy:1960nb}.

It is assumed that we are considering an interacting system for which $H$ and $\bm{K}$ can be expanded in powers of $\frac{1}{m^2}$ as  
\begin{equation}
	\label{generators expansions}
\begin{split}
		H &= Mc^2 + H^{(0)} + H^{(1)} + ..., \\
	\bm{K} &= \bm{K}^{(0)} + \bm{K}^{(1)} + ..., 
\end{split}
\end{equation}
where $M=\sum_\nu^{i} m_\nu$ and the superscripts refer to the order in powers of $\frac{1}{m^2}$. Under this assumption, interactions can be introduced at each order
 as additional terms $V^{(n)}$, $\bm{W}^{(n)}$ to the non-interacting components at the order $n$, respectively $H_0^{(n)}$ and $\bm{K}_0^{(n)}$ ,
\begin{equation}
	\begin{split}
		H^{(n)} &= H_0^{(n)} + V^{(n)}, \\
		\bm{K}^{(n)} &= \bm{K}_0^{(n)} + \bm{W}^{(n)},
	\end{split}
\end{equation}
with both $V^{(n)}$ and $\bm{W}^{(n)}$ depending on the dynamical variables of the system, and not explicitly on time \cite{Foldy:1960nb}.  \\
The commutation relations in Eq. (\ref{poincarè-commrel}) can be consequently expanded in powers of $\frac{1}{m^2}$ and, in principle, they could be solved at each order by mean of a "direct integration". It is conjectured that the obtained solutions represent the most general case for systems of the depicted kind for which the expansion exists. \\




We assume, following \cite{Foldy:1960nb}, that the representation for a relativistic system is chosen so that in the nonrelativistic limit it holds $\bm{W}^{(0)}=0$ and $\bm{K} = M \bm{R} - t\bm{P}$, and  $\bm{J}$ and $\bm{P}$ are those given in Eqs. (\ref{generators}) and (\ref{noninteracting generators}).  

At the first order of the expansions (\ref{generators expansions}), the constraints on $H^{(1)}= H_0^{(1)}+ V^{(1)}$ and $\bm{K}^{(1)}=\bm{K}_0^{(1)} + \bm{W}^{(1)}$ inherited from Poincarè algebra are 
\begin{eqnarray}
		&&[P_i, H^{(1)}] = 0,\nonumber\\
		&&[P_i, K_j^{(1)}] = -i \delta_{ij} \frac{H^{(0)}}{c^2},\nonumber\\
		&&[J_i, H^{(1)}] = 0,\nonumber\\
		&&[J_i, K_j^{(1)}] = i \epsilon_{ijk} K_k^{(1)},\label{poincaré interaction commrel}\\
		&&[H^{(0)}, K_i^{(1)}] + [H^{(1)}, K_i^{(0)}] = 0,\nonumber\\
		&&[K_i^{(0)}, K^{(1)}_j] - [K^{(0)}_j,K^{(1)}_i] = - i \epsilon_{ijk} \frac{J_k}{c^2}.	\nonumber
\end{eqnarray}
The solution of the the above relations allows to identify the relativistic correction $V^{(1)}$ to a 
phenomenological potential $V^{(0)}$.
 
In particular, we focus on the $\bm{P}$-dependent component of such correction, i.e. the boost correction $\delta V(\bm{P})$, which provides the relationship between descriptions of the system in different reference frames.

 An expression for $\delta V$ up to order $\frac{1}{m^2}$ beyond the non-relativistic limit has been derived  by Friar \cite{PhysRevC.12.695}.
 
It is convenient to write it in terms of normalized canonical Jacobi coordinates of  momentum, $\bm{\pi}_{a,b}$ and position $\bm{\rho}_{a,b}$, which are related to the
physical variables $\bm{p}_{1,2,3}$, $\bm{r}_{1,2,3}$  through the change of coordinates
{\small
\begin{equation}
	\label{Jacobi constituents}
 \begin{split}
	\begin{cases}
		\bm{\pi_a} &= \frac{\bm{p_1} - \bm{p_2}}{2}, \\ 
		\bm{\rho_a} &= \bm{r_1} - \bm{r_2};
	\end{cases} 
 \end{split}\quad
 \begin{split}
	 \begin{cases}
		\bm{\pi_b} & = \frac{2}{3}\biggl[ \bm{p_3} -  \frac{\bm{p_1} + \bm{p_2}}{2}\biggr], \\
		\bm{\rho_b} &= \bm{r_3} - \frac{\bm{r_1} + \bm{r_2}}{2};
	\end{cases} \end{split}\end{equation}
}
	{\small
\begin{equation}
	    \begin{cases}
		\bm{R} &= \frac{\bm{r_1} + \bm{r_2} + \bm{r_3}}{3},\nonumber\\
		\bm{P} &= \bm{p_1} + \bm{p_2} + \bm{p_3}.\nonumber
	\end{cases}
\end{equation}
}
Thus, for a system of three particles, $\delta V$ reads
\begin{equation}
	\label{boost correction}
	\delta V(\bm{P}) = - \frac{P^2 V^{(0)}}{2 (3m)^2} - i [\chi_v , H_0] - i [\chi_0, V^{(0)}] , 
\end{equation}
where 
    \begin{align}
	\label{chiv}
	\chi_v (\bm{P}) = &- \frac{1}{6m} \int_{0}^{\bm{P}} \bm{w} \cdot \bm{dP} + H. c., 
 \end{align}
  \begin{align}
		\chi_0 (\bm{P}) =& -\frac{1}{4 (3m)^2} \Biggl[\biggl( \bm{\rho_a} \cdot \bm{P} \bm{\pi_a} \cdot \bm{P}\nonumber\\& + \bm{\rho_b} \cdot \bm{P} \bm{\pi_b} \cdot \bm{P}\biggr) + H. c. \Biggr]  \nonumber\\
		&+	\frac{1}{12 m^2} \Biggl[\biggl( \bm{\rho_a} \cdot \bm{P} \bm{\pi_a} \cdot \bm{\pi_b}  - \frac{1}{2}\bm{\rho_b} \cdot \bm{P} {\bm{\pi_b}}^2 \nonumber\\&+ \frac{2}{3} \bm{\rho_b} \cdot \bm{P} {\bm{\pi_a}}^2 \biggr) + H. c. \Biggr] \nonumber\\
		&- \frac{1}{6 m^2} \biggl[(\bm{s}_1 - \bm{s_2}) \wedge \bm{P} \cdot \bm{\pi_a} \nonumber\\& + (\bm{s}_3 - \frac{\bm{s}_1 + \bm{s}_2}{2} ) \wedge \bm{P} \cdot \bm{\pi_b} \biggr].  \label {chi0 3N jacobi}
\end{align}

We identify $V^{(0)}$ in Eq. (\ref{boost correction}) with the leading order 3N contact interaction parameterized by the LEC $E_0$ in Eq. (\ref{eq:1}). 

The vector $\bm{w}$ in Eq. (\ref{chiv}) is a translationally invariant function of $\bm{P}$ and $\bm{\rho_{a,b}}$, $\bm{\pi_{a,b}}$. It satisfies 
 $\nabla_{\bm{P}} \times \bm{w} = 0$, making the integral in (\ref{chiv}) independent of the path. . As a minimal choice, we set $\bm{w} = 0$,  so that $\chi_v (\bm{P}) = 0$, as in Refs.~\cite{Krajcik:1974nv,Carlson:1993zz,Forest:1995zz,GirlandaSchiavilla};  this corresponds to assuming the existence of an appropriate unitary transformation absorbing $\bm{w}$ \cite{unitarity}.

We evaluate  $\delta V$ between 3N states $\Psi$ and $\Psi'$ as $\bra{\Psi '} \delta V ({\bm P}) \ket{\Psi}$. The details of the calculation can be found in Appendix (\ref{appendix:boost}). 

The result, expressed in terms of variables  $\bm{k_i} = \bm{p_i} - \bm{p_i'}$ and $\bm{Q_i} = \bm{p_i} + \bm{p_i'}$, with $i=1,2,3$, is 
\begin{small}
    \begin{equation}
	\label{delta_V_P relative total momenta}
	\begin{split}
	    \delta V =& - \frac{E_0 }{6m^2} 
 \Biggl[\bm{P^2} 
 + \frac{i}{4} \bm{P} \times (\bm{k_1}-\bm{k_2})\cdot (\bm{\sigma_1}-\bm{\sigma_2})  \\ &
 + \frac{i}{3}\bm{P} \times \biggl(\bm{k_3}- \frac{\bm{k_1}+\bm{k_2}}{2}\biggr) \cdot \biggl(\bm{\sigma_3}- \frac{\bm{\sigma_1}+\bm{\sigma_2}}{2}\biggr)\Biggr].
	\end{split}
\end{equation}
\end{small}
When written in the basis of the 146 3N subleading contact operators $o_{i}$, $i=1,\dots,146$, Eq. (\ref{delta_V_P relative total momenta}) reads
\begin{equation}
	\label{delta_V_P_poincarè}
		\delta V= - \frac{E_0}{48m^2} (-o_{127}-2o_1 + 2 o_{75} - 2 o_{79}). 
\end{equation}
However, when we use the basis of the 18 independent 3N subleading contact operators $O_{1-13}$, $O^{*}_{1-5}$, the same result can be written as 
\begin{equation}
	\label{delta_V_P_poincarè}
		\delta V= \frac{E_0}{24m^2} \left( \frac{2}{3} O^{*}_{1} +O^{*}_{2} \right). 
\end{equation}

Since it is completely characterized by the low energy constant $E_0$, we recognize $\delta {V}({\bm{P}}) $ as the boost correction of our interest, in perfect agreement with Eq.~(\ref{eq:eftcorr}).

\section{Conclusions}
In this paper we derived the relativistic corrections of the leading order 3N contact potential in an arbitrary frame with two different approaches. The first approach, developed within the framework of field theory, involves identifying the operators that contribute to the covariant Lagrangian at the zeroth-order in the low-energy expansion and afterwards performing relativistic reduction; the result is represented by Eqs.~(\ref{eq:drift}) and (\ref{eq:eftcorr}) or, equivalently, by constraints on the LECs $E_i^*$ parametrizing ${\bf P}$-dependent interactions, Eqs. (\ref{lecs dependence on E_0}, \ref{lecs dependence on E_0_2}, \ref{lecs dependence on E_0_3}).  The second approach, formulated within the context of relativistic quantum mechanics, is directly based on fundamental principles of covariance and the constraints that arise from them through Poincaré algebra; the result is Eq. (\ref{delta_V_P_poincarè}).
In both cases it is evident that the $\bm{P}$-dependent boost correction is entirely determined by the interaction at the leading order. The expression for the boost correction is given by
\begin{equation}
	\label{eq:boost}
		\delta V({\bm{P}}) = \frac{E_0}{24m^2} \left( \frac{2}{3} O^{*}_{1} +O^{*}_{2} \right),
\end{equation}
 indicating overall agreement within the two approaches.


The consistency between the two approaches supports the validity of the minimal ansatz $\bm{w}=0$ made in Eq. (\ref{boost correction}). Hence, it is possible to assume the existence of an appropriate unitary transformation absorbing $\bm{w}$ \cite{unitarity}. A similar assumption has been made in the 2N case in Refs. \cite{Forest:1994mw} and \cite{Girlanda:2010zz}; nevertheless, the validity of its extension for systems composed of a $N>2$ nucleons is not straightforward \cite{Krajcik:1974nv}. 

 Furthermore, the results obtained in Eqs (\ref{eq:7})-(\ref{eq:98}) provide a non-trivial check of the Fierz's relations. In fact, if there had been an error in the Fierz's identities, as in Ref. \cite{Girlanda:2011fh} where only 14 nonrelativistic operators were found, the relativistic corrections would have had different expressions for each operator thus precluding the unambiguous determination of the boost correction.

 
It is conjectured that the obtained solutions are the most general for systems of the kind depicted for which the expansion exists. \\
Any relativistic description of a system of interacting particles of finite mass and spin, whether exact or approximate, should fall in this framework.
\begin{appendices}\onecolumn
\section{Non-relativistic expansions }
\label{appendix:expansionsB}
Here we give the non-relativistic expansions of the operators $\tilde{O}_{i}$  defined in Table \ref{tab:operators} in terms of the 146 subleading 3N contact operators $o_i$ and of the six leading order operators $O^{(0)}_i$ listed in Ref. \cite{Girlanda:2011fh}. 
  {\footnotesize
\begin{align}
\label{O_1 nrelexp}
   \tilde{O}_{1} &=  O^{(0)}_1 -\frac{1}{4m^2} \left( 3 o_{75} - \frac{3}{2} o_{127} \right),\\
   \tilde{O}_{2} & = O^{(0)}_3 -\frac{1}{4m^2} \left( o_{76} +o_{77} + o_{78} - o_{128} - \frac{1}{2} o_{129}\right), \\
   \tilde{O}_{3} & = 1 + \frac{1}{4m^2} \left( o_{1} -o_{75} +o_{79} + \frac{1}{2} o_{127} \right), \\
   \tilde{O}_{4} & = O^{(0)}_3+\frac{1}{4m^2} \left( o_{3} +\frac{1}{2} o_{34} - \frac{1}{2}  o_{35} - o_{76} - o_{77} + o_{78}+ o_{81} + \frac{1}{2} o_{128}\right), \\
   \tilde{O}_{5} & = O^{(0)}_3+\frac{1}{4m^2} \left( o_{2} - o_{78} + o_{80} +\frac{1}{2} o_{129}\right), \\
   \tilde{O}_{6} & = O^{(0)}_3+\frac{1}{4m^2} \left(o_{3} -\frac{1}{2} o_{34}+\frac{1}{2} o_{35} -o_{78} + o_{82} + \frac{1}{2}   o_{128} \right), \\
   \tilde{O}_{7} & = O^{(0)}_1 +\frac{1}{4m^2} \left(  o_{1} -2 o_{33} +o_{39}-o_{42}+o_{75}+2 o_{79} + \frac{1}{2} o_{127} \right), \\
   \tilde{O}_{8} & =  O^{(0)}_3+\frac{1}{4m^2} \left( o_{3} -\frac{1}{2} o_{34}-\frac{3}{2}  o_{35} + o_{41}-o_{44}+o_{78}+o_{81}+o_{82} + \frac{1}{2}    o_{128} \right), \\
   \tilde{O}_{9} & = O^{(0)}_3+\frac{1}{4m^2} \left( o_{2}-o_{34}-o_{35}+o_{40}-o_{43}+o_{76}+o_{77}-o_{78}+2 o_{80} +\frac{1}{2}   o_{129} \right) ,\\
   \tilde{O}_{10} & = -O^{(0)}_1-\frac{1}{4m^2} \left(2 o_{1} + o_{75}+2 o_{79} -\frac{1}{2} o_{127} \right), \\
   \tilde{O}_{11} & = -O^{(0)}_3-\frac{1}{4m^2} \left( 2 o_{3}+o_{78}+o_{81}+o_{82} -\frac{1}{2}  o_{129} \right), \\
   \tilde{O}_{12} & = -O^{(0)}_3-\frac{1}{4m^2} \left( o_{2} + o_{3} - \frac{1}{2} o_{34} + \frac{1}{2} o_{35} + o_{76}+o_{77}-o_{78}+o_{80}+o_{82} -\frac{1}{2}   o_{128} \right), \\
   \tilde{O}_{13}  & =  -O^{(0)}_3-\frac{1}{4m^2} \left( o_{2} + o_{3} + \frac{1}{2} o_{34} -\frac{1}{2} o_{35} + o_{78}+o_{80}+o_{81} -\frac{1}{2}   o_{128} \right), \\
   \tilde{O}_{14} & = -O^{(0)}_1-\frac{1}{4m^2} \left( 2 o_{1} + o_{75}+2 o_{79}-\frac{1}{2} o_{127} \right) , \\
   \tilde{O}_{15} & = 
   -O^{(0)}_3-\frac{1}{4m^2} \left( 2 o_{3}+o_{78}+o_{81}+o_{82} - \frac{1}{2}  o_{129} \right) , \\
   \tilde{O}_{16} & =-O^{(0)}_3 - \frac{1}{4m^2} \left( 2 o_{2}+o_{76}+o_{77}-o_{78}+2 o_{80} - o_{128} + \frac{1}{2}o_{129} \right) , \\
   \tilde{O}_{17} & = -O^{(0)}_1- \frac{1}{4m^2} \left( 2 o_{1} + o_{75}+2 o_{79} -\frac{1}{2} o_{127} \right), \\
   \tilde{O}_{18} & = -O^{(0)}_3-\frac{1}{4m^2} \left( o_{2} + o_{3} + \frac{1}{2}  o_{34} - \frac{1}{2} o_{35} + o_{78}+o_{80}+o_{81} - \frac{1}{2}
   o_{128} \right), \\
   \tilde{O}_{19} & = -O^{(0)}_3- \frac{1}{4m^2} \left( 2 o_{3}-o_{34}+o_{35}+o_{76}+o_{77}-o_{78}+2 o_{82} -\frac{1}{2}   o_{129} \right), \\
   \tilde{O}_{20} & = -O^{(0)}_2-\frac{1}{4m^2} \left( o_{4} -\frac{1}{2} o_{36} - \frac{1}{2} o_{39} -\frac{1}{2} o_{45} -\frac{1}{2} o_{49} - o_{79}- o_{83}
   - o_{115}+ o_{130} + \frac{1}{2} o_{134}- o_{137} \right), \\
   \tilde{O}_{21} & = 
   -O^{(0)}_5-\frac{1}{4m^2} \Biggl( o_{6}-\frac{1}{2} o_{38} -\frac{1}{2}  o_{41} -\frac{1}{2} o_{48} - \frac{1}{2} o_{50} -\frac{1}{2} o_{81} -\frac{1}{2} o_{82}- \frac{1}{2} o_{84}- \frac{1}{2} o_{86} + \\ \nonumber
   & \quad - \frac{1}{2}
   o_{116} - \frac{1}{2}o_{117} + \frac{1}{2} o_{132} + \frac{1}{2} o_{133} + \frac{1}{2} o_{135} - \frac{1}{2} o_{139} - \frac{1}{2} o_{140} 
   \Biggr), \\
   \tilde{O}_{22} & = -O^{(0)}_4-\frac{1}{4 m^2} \left( o_{5}- \frac{1}{2} o_{37} - \frac{1}{2}  o_{40} - \frac{1}{2} o_{47} - \frac{1}{2} o_{51} -o_{80}-o_{85}
   - o_{118}+ o_{131} + \frac{1}{2} o_{136}- o_{138} \right), \\
   \tilde{O}_{23} & = 2 O^{(0)}_2+\frac{1}{4m^2} \left( 2 o_{13}-2 o_{17}+2 o_{21}+o_{36}-o_{39}-o_{42}+o_{45}-o_{49}+o_{53}+2 o_{83} -2 o_{115}  +o_{130} \right), \\
   \tilde{O}_{24} & = 2 O^{(0)}_5+\frac{1}{4m^2} \Biggl( 2 o_{14} -2 o_{20}+2 o_{24}+o_{38} -o_{41}-o_{44}+o_{48}-o_{50}+o_{54}-o_{81}+o_{82}-o_{84}+  \\ \nonumber
   & \quad + 2 o_{85}+o_{86} - o_{116} - o_{117} + o_{133} + o_{139} - o_{140} \Biggr), \\
   \tilde{O}_{25} & = 2 O^{(0)}_4+\frac{1}{4m^2} \left( 2 o_{15}-2 o_{19}+2 o_{23}+o_{37}-o_{40}-o_{43}+o_{47}-o_{51}+o_{55}+2 o_{86} -2 o_{118} +    o_{131} \right), \\
   \tilde{O}_{26} & = 2 O^{(0)}_5+\frac{1}{4m^2} \Biggl(2 o_{16}-2 o_{18}+2 o_{22}+o_{38} -o_{41}-o_{44}+o_{46}-o_{52}+o_{56}+o_{81}-o_{82}+3 o_{84} + \\ \nonumber
   & \quad -o_{86} - o_{116} - o_{117} +   o_{132}-o_{139}+o_{140} \Biggr), \\
   \tilde{O}_{27} & = 2 O^{(0)}_2-\frac{1}{4m^2} \left( 2 o_{7} - 2 o_{10} -o_{36}+o_{39}-o_{45}+o_{49}-2 o_{75} + 2 o_{115} - o_{134} - 2 o_{137} \right), \\
   \tilde{O}_{28} & = 2 O^{(0)}_5-\frac{1}{4m^2} \Biggl( 2 o_{9}-    2 o_{12}
   -o_{38}+o_{41}-o_{48}+o_{50}-o_{54}+o_{56}-2 o_{77}+o_{81}-o_{82}+o_{84} + \\ \nonumber
   & \quad -o_{86} + o_{116} + o_{117}- o_{135} -o_{139}-o_{140}\Biggr), \\
   \tilde{O}_{29} & = 2 O^{(0)}_4-\frac{1}{4m^2} \left( 2 o_{8}- 2 o_{11} -o_{37}+o_{40}-o_{47}+o_{51}-2 o_{76} + 2 o_{118}- o_{136} -2 o_{138} \right), \\
   \tilde{O}_{30} & = 2 O^{(0)}_5-\frac{1}{4m^2} \Biggl(2 o_{9}-   2 o_{12}
-o_{38}+o_{41}-o_{46}+o_{52}+o_{54} - o_{56}-2 o_{78}-o_{81}+o_{82}-o_{84}+ \\ \nonumber
   & \quad + o_{86} + o_{116} + o_{117} - o_{135} - o_{139} - o_{140} \Biggr), \\
   \tilde{O}_{31} & = -2 O^{(0)}_2-\frac{1}{4m^2} \left( 2 o_{4}+o_{36}-o_{39}+o_{45}-o_{49}-2 o_{79}-2 o_{83} - 2 o_{115} + 2 o_{130} + o_{134} -2 o_{137} \right), \\
   \tilde{O}_{32} & = -2 O^{(0)}_5-\frac{1}{4m^2} \Biggl( 2 o_{6} +o_{38}-o_{41}+o_{48}-o_{50} -o_{81}-o_{82}-o_{84}-o_{86} - o_{116} - o_{117} + o_{132} + \\ \nonumber
   & \quad+ o_{133} + o_{135} - o_{139} - o_{140} \Biggr), \\
   \tilde{O}_{33} & = -2 O^{(0)}_4-\frac{1}{4m^2} \Biggl( 2 o_{5} + o_{37} - o_{40} + o_{47} - o_{51} - 2 o_{80} - 2 o_{85} - 2 o_{118} + 2 o_{131} + o_{136} - 2 o_{138} \Biggr), \\
   \tilde{O}_{34} & = -2 O^{(0)}_5-\frac{1}{4m^2} \Biggl( 2 o_{6}  +o_{38}-o_{41}+o_{46}-o_{52} -o_{81}-o_{82}-o_{84}-o_{86} - o_{116} - o_{117} + \\ \nonumber
   & \quad+ o_{132} + o_{133} + o_{135} - o_{139} - o_{140} \Biggr), \\
   \tilde{O}_{35} & = 2 O^{(0)}_2-\frac{1}{4m^2} \Biggl( 2 o_{7} - 2 o_{10} + 2 o_{33} - o_{36} - o_{39} +2 o_{42} - o_{45} - o_{49} + 2 o_{53} - 4 o_{75} - 2 o_{79} + \\ \nonumber
   & \quad -2 o_{83} + 2 o_{115} - o_{134} -2 o_{137} \Biggr), \\
   \tilde{O}_{36} & = 2 O^{(0)}_5-\frac{1}{4m^2} \Biggl( 2 o_{9} - 2 o_{12} + 2 o_{35} - o_{38} - o_{41} + 2 o_{44} - o_{48}- o_{50}+ o_{54} + o_{56} - 2 o_{77} - 2 o_{78} + \\ \nonumber
   & \quad - o_{81} - o_{82}- o_{84} - o_{86} + o_{116} + o_{117} - o_{135} - o_{139} - o_{140} \Biggr), \\
   \tilde{O}_{37} & = 2 O^{(0)}_{4}-\frac{1}{4m^2} \Biggl( 2 o_{8} - 2 o_{11} + 2 o_{34} - o_{37} - o_{40} +2 o_{43} - o_{47} - o_{51} +2 o_{55} - 4 o_{76} - 2 o_{80} + \\ \nonumber
   & \quad -2 o_{85} + 2 o_{118} - o_{136} - 2 o_{138} \Biggr), \\
   \tilde{O}_{38} & = O^{(0)}_1+\frac{1}{4 m^2} \left( 2 o_{1} - 2 o_{33} + o_{39} - o_{42} + 3 o_{75} + 3 o_{79} - \frac{1}{2} o_{127} \right), \\
   \tilde{O}_{39} & = O^{(0)}_3+\frac{1}{4m^2} \left(o_{2} + o_{3} - \frac{1}{2} o_{34} - \frac{3}{2} o_{35} + o_{40} - o_{43} + o_{76} + o_{77} + o_{78} + 2 o_{80} + o_{81} - \frac{1}{2} o_{128} \right), \\
   \tilde{O}_{40} & = O^{(0)}_3+\frac{1}{4 m^2} \left(o_{2} + o_{3} - \frac{1}{2} o_{34} - \frac{3}{2} o_{35} + o_{41} - o_{44} + o_{76} + o_{77} + o_{78} + o_{80} + o_{81} + o_{82} - \frac{1}{2} o_{128} \right),\\
        \tilde{O}_{41} & = 
        O^{(0)}_3 + \frac{1}{4m^2} \left( 2 o_{3}-o_{34}-o_{35}+o_{41}-o_{44}+o_{76}+o_{77}+o_{78} + o_{81}+2 o_{82} -\frac{1}{2}  o_{129} \right), \\
        \tilde{O}_{42} & = -O^{(0)}_1-\frac{1}{4m^2} \left(3 o_{1} + 3 o_{75} + 3 o_{79} - \frac{3}{2} o_{127} \right), \\
        \tilde{O}_{43} & = -O^{(0)}_3-\frac{1}{4m^2} \left(o_{2} + 2 o_{3} + o_{76} + o_{77} + o_{78} + o_{80} + o_{81} + o_{82} - o_{128} - \frac{1}{2} o_{129} \right), \\
        \tilde{O}_{44} & = -O^{(0)}_3-\frac{1}{4 m^2} \left(o_{2} + 2 o_{3} + o_{76} + o_{77} + o_{78} + o_{80} + o_{81} + o_{82} - o_{128} - \frac{1}{2} o_{129} \right), \\
        \tilde{O}_{45} & = 
        -O^{(0)}_3-\frac{1}{4m^2} \left( o_{2} + 2 o_{3} + o_{76} + o_{77} + o_{78} + o_{80} + o_{81} + o_{82} - o_{128} - \frac{1}{2} o_{129} \right), \\
        \tilde{O}_{46} & = -O^{(0)}_1-\frac{1}{4m^2} \left( 3 o_{1} + 3 o_{75} + 3 o_{79} - \frac{3}{2} o_{127} \right), \\
        \tilde{O}_{47} & = -O^{(0)}_3-\frac{1}{4m^2} \left(2 o_{2} + o_{3} + \frac{1}{2} o_{34} - \frac{1}{2} o_{35} + o_{76} + o_{77} + o_{78} + 2 o_{80} + o_{81} - \frac{3}{2} o_{128}\right), \\
        \tilde{O}_{48} & = -O^{(0)}_3-\frac{1}{4 m^2} \left(3 o_{3} -\frac{1}{2} o_{34} + \frac{1}{2} o_{35} + o_{76} + o_{77} + o_{78} + o_{81} + 2 o_{82} - \frac{1}{2} o_{128} - o_{129} \right), \\
        \tilde{O}_{49} & = -O^{(0)}_3-\frac{1}{4m^2} \left( o_{2} + 2 o_{3} + o_{76} + o_{77} + o_{78} +o_{80} + o_{81} + o_{82} - o_{128} - \frac{1}{2} o_{129} \right), \\
        \tilde{O}_{50} & = -O^{(0)}_2-\frac{1}{4m^2} \Biggl(o_{4} + o_{21} - \frac{1}{2} o_{36} - \frac{1}{2} o_{39} + \frac{1}{2} o_{42} -\frac{1}{2} o_{45} - \frac{1}{2} o_{49} - \frac{1}{2} o_{53} - o_{79} - o_{83} + \\ \nonumber
   & \quad + o_{115} + o_{119} + \frac{1}{2} o_{130} - o_{137} \Biggr), \\
        \tilde{O}_{51} & = -O^{(0)}_5-\frac{1}{4m^2} \Biggl( o_{6} + o_{22} - \frac{1}{2} o_{38} - \frac{1}{2} o_{41} + \frac{1}{2} o_{44} - \frac{1}{2} o_{48} - \frac{1}{2} o_{50} - \frac{1}{2} o_{56} - \frac{1}{2} o_{81} - \frac{1}{2} o_{82} + \\ \nonumber
   & \quad - \frac{1}{2} o_{84}- \frac{1}{2} o_{86} + \frac{1}{2} o_{116} + \frac{1}{2} o_{117} + o_{120} + \frac{1}{2} o_{132} - \frac{1}{2} o_{139} - \frac{1}{2} o_{140} \Biggr), \\
        \tilde{O}_{52} & = -O^{(0)}_4-\frac{1}{4m^2} \Biggl(o_{5} + o_{23} -\frac{1}{2} o_{37} -\frac{1}{2} o_{40} + \frac{1}{2} o_{43} - \frac{1}{2} o_{47}-\frac{1}{2} o_{51} - \frac{1}{2} o_{55} - o_{80} - o_{85} + \\ \nonumber
   & \quad + o_{118}+ o_{121} + \frac{1}{2} o_{131} - o_{138} \Biggr), \\
        \tilde{O}_{53} & = -O^{(0)}_5-\frac{1}{4m^2} \Biggl( o_{6} + o_{24} -\frac{1}{2} o_{38} - \frac{1}{2} o_{41} + \frac{1}{2} o_{44} - \frac{1}{2} o_{46} - \frac{1}{2} o_{52} - \frac{1}{2} o_{54} - \frac{1}{2} o_{81} - \frac{1}{2} o_{82}+ \\ \nonumber
   & \quad - \frac{1}{2} o_{84} - \frac{1}{2} o_{86} + \frac{1}{2} o_{116} + \frac{1}{2} o_{117} + o_{122} + \frac{1}{2} o_{133} - \frac{1}{2} o_{139} - \frac{1}{2} o_{140} \Biggr), \\
        \tilde{O}_{54} & = 2 O^{(0)}_2+\frac{1}{4m^2} \Biggl( +2 o_{13} -2 o_{17}+2 o_{21}+o_{36} -o_{39}-o_{42}+o_{45}+o_{49}-5 o_{53}+2 o_{83} +2 o_{91} + \\ \nonumber
   & \quad + 2 o_{99} +  2 o_{115}+2 o_{119}+o_{130} \Biggr), \\
        \tilde{O}_{55} & = 2 O^{(0)}_5+\frac{1}{4m^2} \Biggl( 2 o_{14} -2 o_{20}+2 o_{24}+o_{38}-o_{41}-o_{44}+o_{48}-o_{50}+2 o_{51}-o_{54} -2 o_{55} + \\ \nonumber
   & \quad -2 o_{56}-o_{81}+o_{82}-o_{84}+2 o_{85}+o_{86}+2 o_{92}
 +       2 o_{102}+o_{116}+o_{117}+2 o_{122}+o_{133}+o_{139} -o_{140} \Biggr), \\
        \tilde{O}_{56} & = 
        2 O^{(0)}_4+\frac{1}{4m^2} \Biggl( 2 o_{15}-2 o_{19}+2 o_{23}+o_{37} -o_{40}-o_{43}+o_{47}-o_{51}+2 o_{52}-3 o_{55} -2 o_{56} + \\ \nonumber
   & \quad + 2 o_{86} + 2 o_{93}+
        2 o_{101}+2 o_{118}+2 o_{121}+o_{131} \Biggr), \\
        \tilde{O}_{57} & = 
        2 O^{(0)}_5+\frac{1}{4m^2} \Biggl( 2 o_{16} -2 o_{18}+2 o_{22}+o_{38}-o_{41}-o_{44}+o_{46}+2 o_{50}-o_{52}-4 o_{54}-o_{56} + \\ \nonumber
   & \quad + o_{81}-o_{82}+3 o_{84}-o_{86}+2 o_{94}
+  2 o_{100}+o_{116}+o_{117}+2 o_{120}+o_{132}-o_{139}+o_{140}\Biggr), \\
\tilde{O}_{58} & = -2 O^{(0)}_2-\frac{1}{4m^2} \left( 2 o_{13} -2 o_{17}+4 o_{21}+o_{36} -o_{39} + o_{45}-o_{49}+2 o_{83} +
2 o_{115}+2 o_{119}-o_{134} \right), \\
\tilde{O}_{59} & = -2 O^{(0)}_5-\frac{1}{4m^2} \Biggl( 2 o_{14} -2 o_{20}+  2 o_{22}+2 o_{24}+o_{38}-o_{41}+o_{48}-o_{50}+o_{54}-o_{56}-o_{81} + \\ \nonumber
   & \quad + o_{82}-o_{84}+2 o_{85}+o_{86} +
o_{116}+o_{117}+2 o_{120}-o_{135}+o_{139}-o_{140} \Biggr), \\
\tilde{O}_{60} & = -2 O^{(0)}_4-\frac{1}{4m^2} \left( 2 o_{15}-2 o_{19}+4 o_{23}+o_{37}-o_{40} + o_{47}-o_{51}+2 o_{86}+ 2 o_{118}+2 o_{121}-o_{136}\right), \\
        \tilde{O}_{61} & = 
        -2 O^{(0)}_5-\frac{1}{4m^2} \Biggl(2 o_{16} -2 o_{18}  +2 o_{22}+2 o_{24}+o_{38}-o_{41}+o_{46}-o_{52}-o_{54}+o_{56} +o_{81} + \\ \nonumber
   & \quad-o_{82}+3 o_{84}-o_{86} +
        o_{116}+o_{117}+2 o_{122}-o_{135}-o_{139}+o_{140} \Biggr), \\
        \tilde{O}_{62} & = -2 O^{(0)}_2+\frac{1}{4m^2} \Biggl( 2 o_{7} - 2 o_{10} - 2 o_{21} - o_{36} + o_{39} - o_{42} - o_{45} + o_{49} + o_{53} - 2 o_{75} - 2 o_{115} + \\ \nonumber
   & \quad - 2 o_{119} + o_{130} - 2 o_{137} \Biggr), \\
        \tilde{O}_{63} & = 
        -2 O^{(0)}_5+\frac{1}{4m^2} \Biggl( 2 o_{9} - 2 o_{12} - 2 o_{22} - o_{38} + o_{41} - o_{44} - o_{48} +o_{50} - o_{54} + 2 o_{56} - 2 o_{77} + \\ \nonumber
   & \quad + o_{81} - o_{82} + o_{84} - o_{86} - o_{116} - o_{117} - 2 o_{120} + o_{133} - o_{139} - o_{140} \Biggr), \\
        \tilde{O}_{64} & = 
        -2 O^{(0)}_4+\frac{1}{4m^2} \Biggl(2 o_{8} - 2 o_{11} - 2 o_{23} - o_{37} + o_{40} - o_{43} - o_{47} + o_{51} + o_{55} - 2 o_{76} - 2 o_{118} + \\ \nonumber
   & \quad - 2 o_{121} + o_{131} - 2 o_{138} \Biggr),\\
        \tilde{O}_{65} & = 
        -2 O^{(0)}_5+\frac{1}{4m^2} \Biggl( 2 o_{9} - 2 o_{12} - 2 o_{24} - o_{38} + o_{41} - o_{44} - o_{46} + o_{52} + 2 o_{54} - o_{56} - 2 o_{78} + \\ \nonumber
   & \quad - o_{81} + o_{82} -o_{84} + o_{86} - o_{116} - o_{117} - 2 o_{122} + o_{132} - o_{139} - o_{140} \Biggr), \\
        \tilde{O}_{66} & = 2 O^{(0)}_2+\frac{1}{4m^2} \Biggl( 2 o_{4} +2 o_{21}+o_{36}-o_{39} +o_{42}+o_{45}-o_{49}-o_{53}-2 o_{79} - 2 o_{83} +        2 o_{115} + \\ \nonumber
   & \quad +2 o_{119}+o_{130}-2 o_{137} \Biggr), \\
        \tilde{O}_{67} & = 
        2 O^{(0)}_5+\frac{1}{4m^2} \Biggl(2 o_{6}+2 o_{22}+ o_{38}-o_{41}+o_{44}+o_{48}-o_{50}-o_{56} -o_{81}-o_{82}-o_{84} + \\ \nonumber
   & \quad  -o_{86} +
        o_{116}+o_{117}+2 o_{120}+o_{132}-o_{139}-o_{140} \Biggr), \\
        \tilde{O}_{68} & = 
        2 O^{(0)}_4+\frac{1}{4m^2} \Biggl( 2 o_{5} +2 o_{23}+o_{37}-o_{40} + o_{43}+o_{47}-o_{51}-o_{55}-2 o_{80} -2 o_{85} + 
        2 o_{118}+ \\ \nonumber
   & \quad  +2 o_{121}+o_{131}-2 o_{138}\Biggr), \\
        \tilde{O}_{69} & = 
        2 O^{(0)}_5+\frac{1}{4m^2} \Biggl(2 o_{6} +2 o_{24}+o_{38}-o_{41}+o_{44}+o_{46}-o_{52}-o_{54}-o_{81}-o_{82}-o_{84}-o_{86} + \\ \nonumber
   & \quad  +       o_{116}+o_{117}+2 o_{122}+o_{133}-o_{139}-o_{140} \Biggr), \\
        \tilde{O}_{70} & = -2 O^{(0)}_2-\frac{1}{4m^2} \left( 2 o_{13} -2 o_{17}+4 o_{21}+o_{36} -o_{39}+o_{45}-o_{49}+2 o_{83} +
        2 o_{115}+2 o_{119}-o_{134}\right),\\
        \tilde{O}_{71} & = 
        -2 O^{(0)}_5-\frac{1}{4m^2} \Biggl(2 o_{14} -2 o_{20}+4 o_{24}+o_{38}-o_{41}+o_{48}-o_{50}-o_{81}+o_{82} -o_{84}+2 o_{85} + \\ \nonumber
   & \quad  +o_{86} +
        o_{116}+o_{117}+2 o_{122}-o_{132}+o_{133}-o_{135}+o_{139}-o_{140} \Biggr), \\
        \tilde{O}_{72} & = 
        -2 O^{(0)}_4-\frac{1}{4m^2} \left( 2 o_{15}-2 o_{19}+4 o_{23}+o_{37}  -o_{40}+o_{47}-o_{51}+2 o_{86} +        2 o_{118}+2 o_{121}-o_{136} \right), \\
        \tilde{O}_{73} & = 
        -2 O^{(0)}_5-\frac{1}{4m^2} \Biggl(2 o_{16}-2 o_{18}+4 o_{22}+o_{38}-o_{41}+o_{46}-o_{52}+o_{81}-o_{82}+ 3 o_{84}-o_{86} + \\ \nonumber
   & \quad  +       o_{116}+o_{117}+2 o_{120}+o_{132}-o_{133}-o_{135}-o_{139} + o_{140} \Biggr), \\
        \tilde{O}_{74} & = -2 O^{(0)}_2+\frac{1}{4m^2} \Biggl(2 o_{7} - 2 o_{10} - 2 o_{21} - o_{36} + o_{39} - o_{42} - o_{45} + o_{49} +o_{53} -2 o_{75} -2 o_{115} + \\ \nonumber
   & \quad  -2 o_{119} + o_{130} -2 o_{137} \Biggr), \\
        \tilde{O}_{75} & = 
        -2 O^{(0)}_5+\frac{1}{4m^2} \Biggl(2 o_{9} -2 o_{12} -2 o_{24} - o_{38} + o_{41} -o_{44} - o_{48} + o_{50} +o_{56} -2 o_{77} + o_{81} + \\ \nonumber
   & \quad  - o_{82} + o_{84} -o_{86} -o_{116} -o_{117} -2 o_{122} + o_{132} -o_{139} -o_{140} \Biggr), \\
        \tilde{O}_{76} & = 
        -2 O^{(0)}_4+\frac{1}{4m^2} \Biggl(2 o_{8} - 2 o_{11} - 2 o_{23} - o_{37} + o_{40} - o_{43} - o_{47} + o_{51} + o_{55} - 2 o_{76} -2 o_{118} + \\ \nonumber
   & \quad  -2 o_{121} + o_{131} -2 o_{138} \Biggr), \\
        \tilde{O}_{77} & = 
        -2 O^{(0)}_5+ \frac{1}{4m^2} \Biggl(2 o_{9} -2 o_{12} - 2 o_{22} -o_{38} + o_{41} -o_{44} - o_{46} + o_{52} + o_{54} -2 o_{78} - o_{81} + \\ \nonumber
   & \quad  + o_{82} - o_{84} + o_{86} - o_{116} - o_{117} -2 o_{120} + o_{133} - o_{139} -o_{140} \Biggr), \\
        \tilde{O}_{78} & = 2 O^{(0)}_2+\frac{1}{4m^2} \Biggl(2 o_{4} +2 o_{21}+o_{36}-o_{39} +o_{42}+o_{45}-o_{49}-o_{53}-2 o_{79}-2 o_{83} +
        2 o_{115} + \\ \nonumber
   & \quad  +2 o_{119}+o_{130}-2 o_{137}\Biggr), \\
        \tilde{O}_{79} & = 
        2 O^{(0)}_5+\frac{1}{4m^2} \Biggl( 2 o_{6}+2 o_{24}+o_{38}-o_{41}+o_{44}+o_{48}-o_{50}-o_{54}-o_{81}-o_{82}-o_{84}-o_{86} + \\ \nonumber
   & \quad  + 
        o_{116}+o_{117}+2 o_{122}+o_{133}-o_{139}-o_{140} \Biggr), \\
        \tilde{O}_{80} & = 
        2 O^{(0)}_4+\frac{1}{4m^2} \Biggl(2 o_{5}+2 o_{23}+o_{37}-o_{40} + o_{43}+o_{47}-o_{51}-o_{55}-2 o_{80}-2 o_{85}+
        2 o_{118} + \\ \nonumber
   & \quad +2 o_{121}+o_{131}-2 o_{138}\Biggr), \\
        \tilde{O}_{81} & = 
        2 O^{(0)}_5+\frac{1}{4m^2} \Biggl(2 o_{6} +2 o_{22}+o_{38}-o_{41}+o_{44}+o_{46}-o_{52}-o_{56}-o_{81}-o_{82}-o_{84}-o_{86} + \\ \nonumber
   & \quad +
        o_{116}+o_{117}+2 o_{120}+o_{132}-o_{139}-o_{140}\Biggr), \\
        \tilde{O}_{82} & = 2 O^{(0)}_2-\frac{1}{4m^2} \Biggl(2 o_{7} -2 o_{10} - 2 o_{21} + 2 o_{33} - o_{36} - o_{39} + o_{42} - o_{45}- o_{49} + 3 o_{53} - 4 o_{75}+ \\ \nonumber
   & \quad   -2 o_{79} - 2 o_{83} - 2 o_{115} -2 o_{119}  + o_{130} - 2 o_{137} \Biggr), \\
        \tilde{O}_{83} & = 
        2 O^{(0)}_5-\frac{1}{4m^2} \Biggl(2 o_{9} - 2 o_{12} - 2 o_{22} + 2 o_{35} - o_{38} - o_{41} + o_{44} - o_{48} - o_{50} + o_{54} + 2 o_{56} + \\ \nonumber
   & \quad  - 2 o_{77} -2 o_{78} - o_{81} - o_{82} - o_{84} - o_{86} - o_{116} - o_{117} - 2 o_{120} + o_{133} - o_{139} - o_{140} \Biggr), \\
        \tilde{O}_{84} & = 
        2 O^{(0)}_4-\frac{1}{4m^2} \Biggl(2 o_{8} - 2 o_{11} - 2 o_{23} + 2 o_{34} - o_{37} - o_{40} + o_{43}-o_{47} - o_{51} + 3 o_{55} - 4 o_{76} + \\ \nonumber
   & \quad  -2 o_{80} -2 o_{85} -2 o_{118} -2 o_{121} + o_{131} -2 o_{138} \Biggr), \\
        \tilde{O}_{85} & = 
        2 O^{(0)}_5-\frac{1}{4m^2} \Biggl(2 o_{9} -2 o_{12} - 2o_{24} + 2 o_{35} - o_{38}- o_{41} + o_{44} - o_{46} -o_{52} +2 o_{54}+ o_{56} + \\ \nonumber
   & \quad  -2 o_{77} -2 o_{78} - o_{81} - o_{82} - o_{84} - o_{86} - o_{116} - o_{117} -2 o_{122} + o_{132} - o_{139} - o_{140} \Biggr), \\
        \tilde{O}_{86} & = O^{(0)}_6-\frac{1}{4m^2} \Biggl(2 o_{29} - 2 o_{31} + 2 o_{32} - o_{59} + \frac{3}{2} o_{60} -\frac{3}{2} o_{61} - \frac{3}{2} o_{63} + \frac{3}{2} o_{64} + 3 o_{72} - 3 o_{73} + \\ \nonumber
   & \quad  - o_{144} + \frac{1}{2} o_{145} \Biggr), \\
          \tilde{O}_{87} & = 
          O^{(0)}_6-\frac{1}{4m^2} \Biggl(2 o_{29} - 2 o_{31} + 2 o_{32} + o_{59} - \frac{1}{2} o_{60} -\frac{3}{2} o_{61} + \frac{1}{2} o_{63} -\frac{1}{2} o_{64} + 2 o_{69} + o_{72} + \\ \nonumber
   & \quad  - o_{73} - o_{144} + \frac{1}{2} o_{145} \Biggr), \\
        \tilde{O}_{88} & = 
        O^{(0)}_6+\frac{1}{4m^2} \Biggl( 3 o_{30} + 3 o_{31} - o_{58} - o_{59} + \frac{3}{2} o_{60} + \frac{1}{2} o_{61} - \frac{3}{2} o_{63} + \frac{3}{2} o_{64} + 4 o_{70} - 4 o_{71} + \\ \nonumber
   & \quad  + o_{72} - o_{73} + \frac{3}{2} o_{145} \Biggr), \\
        \tilde{O}_{89} & = 
        O^{(0)}_6-\frac{1}{4m^2} \Biggl(2 o_{29} - o_{30} -2 o_{31} + 2 o_{32} + o_{58} + o_{59} - \frac{3}{2} o_{60} - \frac{1}{2} o_{61}+ \frac{3}{2} o_{63} - \frac{3}{2} o_{64}  + \\ \nonumber
   & \quad  + o_{69} -2 o_{70} + o_{71} - \frac{1}{2} o_{145} + o_{146} \Biggr), \\
        \tilde{O}_{90} & = 
        -2 O^{(0)}_6-\frac{1}{4m^2} \Biggl(2 o_{27}-2 o_{28}-2 o_{29}+2 o_{30}+4 o_{31} -4 o_{32}+o_{60}+3 o_{61}-2 o_{62}-3 o_{63} + \\ \nonumber
   & \quad  +3 o_{64}-4 o_{69}+4 o_{70}+2 o_{72}-2 o_{73}+    o_{144}+2 o_{146}  \Biggr), \\
          \tilde{O}_{91} & = 
          -2 O^{(0)}_6-\frac{1}{4m^2} \Biggl(2 o_{27}-2 o_{28}-2 o_{29}+4 o_{31}-4 o_{32}+o_{60} +3 o_{61}-2 o_{62}-3 o_{63}+3 o_{64} + \\ \nonumber
   & \quad  -4 o_{69}+2 o_{70} +      o_{144} \Biggr), \\
          \tilde{O}_{92} & = -O^{(0)}_6-\frac{1}{4m^2} \Biggl( 3 o_{30} + 3 o_{31} - 3 o_{58} - 3 o_{59} + \frac{3}{2} o_{60} - \frac{3}{2} o_{61} - \frac{3}{2} o_{63} + \frac{3}{2} o_{64} + 6 o_{70} + \\ \nonumber
   & \quad  - 6 o_{71} + 3 o_{72} - 3 o_{73} + \frac{3}{2} o_{145} \Biggr).
\end{align}
     }
\section{Reduction to the minimal basis}
\label{appendix:expansionsO}
Here we use the Fierz's relations among the subleading $3N$ contact operators derived in Ref.~\cite{Girlanda:2011fh} to rewrite the non-relativistic expansions of the 92 relativistic operators in terms of the minimal basis of operators appearing in Eq.~(\ref{eq:1}). Notice that, as a further consequence of Fierz' identities, the six leading order operators $O^{(0)}_{i=1,...,6}$ are all proportional to the operator $O_0$, which is the identity operator in spin and isospin space, 
\begin{equation}
  O_0 \equiv O^{(0)}_1= - O^{(0)}_2= - O^{(0)}_3 = -\frac{1}{3} O^{(0)}_4= \frac{1}{3} O^{(0)}_5=-\frac{1}{12} O^{(0)}_6.
  \end{equation}
{\footnotesize 
\begin{align}
    \tilde{O}_1 & = O^{(0)}_{\bf{P}} - \frac{1}{4m^2}\left( O_1 + \frac{1}{2} O_2 + \frac{1}{2} O_3 + \frac{1}{2} O_4 + 2 O_7 - \frac{3}{4} O_9 - \frac{3}{4} O_{10} + \frac{3}{4} O_{11} + \frac{3}{4} O_{12} \right),\label{eq:7} \\
    \tilde{O}_2 &= - O^{(0)}_{\bf{P}} + \frac{1}{4m^2}\Biggl(
O_1 -\frac{3}{2} O_2 + \frac{5}{2} O_3 + \frac{7}{6} O_4+2 O_5 + \frac{2}{3} O_6 + 2 O_7 + \frac{9}{4} O_9 + \frac{1}{4} O_{10} + \\ \nonumber
   & \quad  + \frac{3}{4} O_{11} + \frac{3}{4} O_{12} - 2 O_{13} \Biggr), \\    %
\tilde{O}_3 &= O^{(0)}_{\bf{P}} \\ 
\tilde{O}_4 &= -O^{(0)}_{\bf{P}} - \frac{1}{4m^2} \left(
2 O_1 + O_2 + O_3 - 8 O_7 - 4 O_8 -\frac{3}{2} O_9 -\frac{1}{2} O_{10} + \frac{9}{2} O_{11} + \frac{3}{2} O_{12} + 2 O_{13} \right), \\
\tilde{O}_5 &= -O^{(0)}_{\bf{P}} + \frac{1}{4m^2} \left( O_1 - 2 O_2 + 3 O_3 + \frac{1}{3} O_4 + O_5 + \frac{1}{3} O_6 + 4 O_7 -2 O_{13} \right), \\
\tilde{O}_6 &= -O^{(0)}_{\bf{P}}
+ \frac{1}{m^2}\bigl( 
- \frac{5}{8} O_9 -O_8-3O_7-  \frac{1}{12} O_6 - \frac{1}{4} O_5-\frac{1}{12}O_4 - \frac{1}{2} O_3 +  \\ \nonumber 
&\qquad + \frac{3}{4} O_2 +O_{13}+\frac{5}{8} O_{12}+\frac{11}{8}O_{11}-\frac{3}{8}O_{10}+\frac{1}{4}O_1 \bigr), \\
\tilde{O}_7 &= O^{(0)}_{\bf{P}} + \frac{1}{m^2}\bigl( -\frac{7}{16}O_9+\frac{1}{2}O_7-\frac{1}{12}O_6-\frac{1}{4}O_5-\frac{1}{12}O_4-\frac{1}{16}O_{12}-\frac{1}{16}O_{11}-\frac{3}{16}O_{10}+\frac{1}{8}O_{1}\bigr), \\
\tilde{O}_8 &= -O^{(0)}_{\bf{P}} + \frac{1}{m^2}\bigl(
-\frac{13}{16}O_{9}-\frac{3}{2}O_{7}-\frac{1}{4}O_{6}-\frac{3}{4}O_{5}-\frac{3}{8}O_{4}-\frac{11}{8}O_{3}+\frac{9}{8}O_{2}+\frac{5}{4}O_{13}+\frac{1}{16}O_{12}+ \\ \nonumber 
&\qquad+ \frac{1}{16}O_{11}-\frac{5}{16}O_{10}-\frac{1}{2}O_{1}\bigr), \\
\tilde{O}_9 &=-O^{(0)}_{\bf{P}} + \frac{1}{m^2}\bigl( 
-\frac{1}{16}O_{9}+\frac{3}{2}O_{7}+\frac{1}{12}O_6+\frac{1}{4}O_5+\frac{1}{3}O_{4}+\frac{3}{4}O_{3}-\frac{1}{4}O_{2}-\frac{1}{2}O_{13}+\frac{1}{16}O_{12}+ \\ \nonumber 
&\qquad+\frac{1}{16}O_{11}+\frac{3}{16}O_{10}+\frac{5}{8}O_{1}\bigr), \\
\tilde{O}_{10} &= -O^{(0)}_{\bf{P}} + \frac{1}{m^2}\bigl(
+\frac{1}{16}O_{9}-\frac{1}{2}O_7-\frac{1}{8}O_{4}-\frac{1}{8}O_{3}-\frac{1}{8}O_{2}-\frac{1}{16}O_{12}-\frac{1}{16}O_{11}+\frac{1}{16}O_{10}-\frac{1}{4}O_1\bigr), \\
\tilde{O}_{11} &= O^{(0)}_{\bf{P}} + \frac{1}{m^2}\bigl(
+\frac{13}{16}O_{9}+\frac{3}{2}O_{7}+\frac{1}{4}O_{6}+\frac{3}{4}O_{5}+\frac{3}{8}O_{4}+\frac{11}{8}O_{3}-\frac{7}{8}O_{2}-O_{13}-\frac{1}{16}O_{12}+ \\ \nonumber 
&\qquad -\frac{1}{16}O_{11}+\frac{5}{16}O_{10}+\frac{1}{2}O_{1}\bigr), \\
\tilde{O}_{12} &= O^{(0)}_{\bf{P}} + \frac{1}{m^2}\bigl(
+\frac{19}{16}O_{9}+O_8+\frac{5}{2}O_{7}+\frac{1}{6}O_{6}+\frac{1}{2}O_{5}+\frac{7}{24}O_{4}+\frac{3}{8}O_{3}-\frac{5}{8}O_{2}-O_{13}-\frac{7}{16}O_{12}+ \\ \nonumber 
&\qquad-\frac{19}{16}O_{11}+\frac{7}{16}O_{10}-\frac{1}{4}O_1\bigr), \\
\tilde{O}_{13} &= O^{(0)}_{\bf{P}} + \frac{1}{m^2}\bigl(
+\frac{1}{16}O_{9}-O_8-\frac{5}{2}O_{7}+\frac{1}{12}O_6+\frac{1}{4}O_5+\frac{5}{24}O_{4}+\frac{1}{8}O_{3}+\frac{3}{8}O_{2}+\frac{1}{2}O_{13}+ \\ \nonumber 
&\qquad +\frac{11}{16}O_{12}+\frac{23}{16}O_{11}-\frac{3}{16}O_{10}+\frac{1}{2}O_{1}\bigr), \\
\tilde{O}_{14} &= -O^{(0)}_{\bf{P}} + \frac{1}{m^2}\bigl(
+\frac{1}{16}O_{9}-\frac{1}{2}O_7-\frac{1}{8}O_{4}-\frac{1}{8}O_{3}-\frac{1}{8}O_{2}-\frac{1}{16}O_{12}-\frac{1}{16}O_{11}+\frac{1}{16}O_{10}-\frac{1}{4}O_1\bigr), \\
\tilde{O}_{15} &=O^{(0)}_{\bf{P}} + \frac{1}{m^2}\bigl(
+\frac{13}{16}O_{9}+\frac{3}{2}O_{7}+\frac{1}{4}O_{6}+\frac{3}{4}O_{5}+\frac{3}{8}O_{4}+\frac{11}{8}O_{3}-\frac{7}{8}O_{2}-O_{13}-\frac{1}{16}O_{12}+ \\ \nonumber 
&\qquad-\frac{1}{16}O_{11}+\frac{5}{16}O_{10}+\frac{1}{2}O_{1}\bigr), \\
\tilde{O}_{16} &= O^{(0)}_{\bf{P}} + \frac{1}{m^2}\bigl(
+\frac{7}{16}O_9-\frac{3}{2}O_{7}+\frac{1}{8}O_{4}-\frac{7}{8}O_{3}+\frac{5}{8}O_{2}+\frac{1}{2}O_{13}+\frac{5}{16}O_{12}+\frac{5}{16}O_{11}-\frac{1}{16}O_{10}-\frac{1}{4}O_1\bigr), \\
\tilde{O}_{17} &= -O^{(0)}_{\bf{P}} + \frac{1}{m^2}\bigl(
+\frac{1}{16}O_{9}-\frac{1}{2}O_7-\frac{1}{8}O_{4}-\frac{1}{8}O_{3}-\frac{1}{8}O_{2}-\frac{1}{16}O_{12}-\frac{1}{16}O_{11}+\frac{1}{16}O_{10}-\frac{1}{4}O_1\bigr), \\
\tilde{O}_{18} &= O^{(0)}_{\bf{P}} + \frac{1}{m^2}\bigl(
+\frac{1}{16}O_{9}-O_8-\frac{5}{2}O_{7}+\frac{1}{12}O_6+\frac{1}{4}O_5+\frac{5}{24}O_{4}+\frac{1}{8}O_{3}+\frac{3}{8}O_{2}+\frac{1}{2}O_{13}+\frac{11}{16}O_{12}+ \\ \nonumber 
&\qquad+\frac{23}{16}O_{11}-\frac{3}{16}O_{10}+\frac{1}{2}O_{1}\bigr), \\
\tilde{O}_{19} &= O^{(0)}_{\bf{P}} + \frac{1}{m^2}\bigl(
+\frac{31}{16}O_{9}+2O_8+\frac{13}{2}O_{7}+\frac{1}{3}O_{6}+O_5+\frac{11}{24}O_{4}+\frac{13}{8}O_{3}-\frac{15}{8}O_{2}-\frac{5}{2}O_{13}-\frac{19}{16}O_{12}+ \\ \nonumber 
&\qquad-\frac{43}{16}O_{11}+\frac{15}{16}O_{10}-\frac{1}{4}O_1\bigr), \\
\tilde{O}_{20} &= O^{(0)}_{\bf{P}} + \frac{1}{m^2}\bigl(
+\frac{9}{16}O_{9}+\frac{1}{2}O_7+\frac{1}{12}O_6+\frac{1}{4}O_5+\frac{1}{12}O_4+\frac{1}{2}O_{3}-\frac{1}{2}O_{2}-\frac{1}{2}O_{13}-\frac{1}{16}O_{12}+ \\ \nonumber 
&\qquad-\frac{1}{16}O_{11}+\frac{5}{16}O_{10}+\frac{1}{8}O_{1}\bigr), \\
\tilde{O}_{21} &= -3 O^{(0)}_{\bf{P}} + \frac{1}{m^2}\bigl(
+\frac{9}{16}O_{9}-\frac{1}{2}O_7+\frac{1}{4}O_{6}+\frac{3}{4}O_{5}+\frac{3}{8}O_{4}+\frac{3}{8}O_{3}-\frac{1}{8}O_{2}-\frac{1}{4}O_{13}+\frac{3}{16}O_{12}+ \\ \nonumber 
&\qquad+\frac{3}{16}O_{11}+\frac{1}{16}O_{10}\bigr), \\
      %
\tilde{O}_{22} &= O^{(0)}_{\bf{P}} + \frac{1}{m^2}\bigl(
+\frac{3}{16}O_{9}-\frac{1}{2}O_7-\frac{1}{12}O_6-\frac{1}{4}O_5-\frac{1}{3}O_{4}-\frac{1}{4}O_{3}-\frac{1}{4}O_{2}-\frac{3}{16}O_{12}-\frac{3}{16}O_{11}+ \\ \nonumber 
&\qquad-\frac{1}{16}O_{10}-\frac{3}{8}O_{1}\bigr), \\
\tilde{O}_{23} &= -2O^{(0)}_{\bf{P}} + \frac{1}{m^2}\bigl(
+\frac{5}{4}O_{9}+2O_8+6O_7+\frac{1}{6}O_{6}+\frac{1}{2}O_{5}+\frac{1}{6}O_{4}+O_3-\frac{3}{2}O_{2}-2O_{13}-\frac{5}{4}O_{12}+ \\ \nonumber 
&\qquad-\frac{11}{4}O_{11}+\frac{3}{4}O_{10}-\frac{1}{2}O_{1}\bigr), \\
\tilde{O}_{24} &= 6 O^{(0)}_{\bf{P}} + \frac{1}{m^2}\bigl( -\frac{3}{2}O_{9}-2O_8-8O_7-\frac{1}{3}O_{6}-\frac{1}{3}O_{4}-\frac{3}{2}O_{3}+\frac{3}{2}O_{2}+2O_{13}+\frac{3}{2}O_{12}+3O_{11}-2O_{10}\bigr), \\
\tilde{O}_{25} &= -6 O^{(0)}_{\bf{P}}, \\
\tilde{O}_{26} &= 6 O^{(0)}_{\bf{P}} + \frac{1}{m^2}\bigl( -\frac{9}{4}O_{9}-4O_8-10O_7-\frac{1}{6}O_{6}-\frac{3}{2}O_{5}-\frac{1}{6}O_{4}-\frac{3}{2}O_{3}+3O_2+4O_{13}+\frac{9}{4}O_{12}++ \\ \nonumber 
&\qquad \frac{21}{4}O_{11}-\frac{1}{4}O_{10}+\frac{3}{2}O_{1}\bigr), \\
\tilde{O}_{27} &=-2 O^{(0)}_{\bf{P}} + \frac{1}{m^2}\bigl( -\frac{3}{4}O_{9}-\frac{1}{3}O_{6}-\frac{5}{6}O_{5}-\frac{1}{12}O_4-\frac{13}{12}O_{3}+\frac{5}{4}O_{2}+O_{13}+\frac{1}{4}O_{12}+\frac{1}{4}O_{11}-\frac{1}{4}O_{10}+\frac{1}{4}O_1\bigr), \\
\tilde{O}_{28} &= 6 O^{(0)}_{\bf{P}} + \frac{1}{m^2}\bigl(
-3O_9-2O_7-\frac{1}{3}O_{6}-\frac{1}{2}O_{5}-\frac{7}{12}O_{4}-\frac{5}{4}O_{3}+\frac{1}{4}O_{2}+\frac{1}{2}O_{13}-O_{10}-\frac{3}{4}O_{1}\bigr), \\
\tilde{O}_{29} &= -6 O^{(0)}_{\bf{P}}+ \frac{1}{m^2}\bigl( 
+2O_7-\frac{1}{2}O_{5}+\frac{1}{4}O_{4}+\frac{1}{4}O_{3}+\frac{1}{4}O_{2}-\frac{1}{2}O_{10}+\frac{3}{4}O_{1}\bigr), \\
\tilde{O}_{30} &= 6 O^{(0)}_{\bf{P}} + \frac{1}{m^2}\bigl(
-\frac{9}{4}O_{9}-\frac{1}{3}O_{6}-\frac{3}{2}O_{5}-\frac{13}{12}O_{4}-\frac{3}{4}O_{3}+\frac{3}{4}O_{2}+\frac{3}{2}O_{13}-\frac{3}{4}O_{12}-\frac{3}{4}O_{11}+ \\ \nonumber 
&\qquad-\frac{1}{4}O_{10}-\frac{3}{4}O_{1}\bigr), \\
\tilde{O}_{31} &= 2 O^{(0)}_{\bf{P}} + \frac{1}{m^2}\bigl(
+\frac{9}{8}O_{9}+O_7+\frac{1}{6}O_{6}+\frac{2}{3}O_{5}+\frac{1}{6}O_{4}+\frac{7}{6}O_3-O_2-O_{13}-\frac{1}{8}O_{12}+ \\ \nonumber 
&\qquad-\frac{1}{8}O_{11}+\frac{5}{8}O_{8}+\frac{1}{4}O_1\bigr), \\
\tilde{O}_{32} &= -6 O^{(0)}_{\bf{P}} + \frac{1}{m^2}\bigl(
+\frac{9}{8}O_{9}-O_7+\frac{1}{2}O_{6}+O_5+\frac{3}{4}O_{4}+\frac{1}{4}O_{3}-\frac{1}{4}O_{2}-\frac{1}{2}O_{13}+\frac{3}{8}O_{12}+ \\ \nonumber 
&\qquad+\frac{3}{8}O_{11}+\frac{1}{8}O_{10}\bigr), \\
\tilde{O}_{33} &= 6 O^{(0)}_{\bf{P}} + \frac{1}{m^2}\bigl(
+\frac{3}{8}O_{9}-O_7-\frac{1}{2}O_{5}-\frac{1}{2}O_{4}-\frac{1}{2}O_{3}-\frac{1}{2}O_{2}-\frac{3}{8}O_{12}-\frac{3}{8}O_{11}-\frac{1}{8}O_{10}-\frac{3}{4}O_{1}\bigr), \\
\tilde{O}_{34} &= -6 O^{(0)}_{\bf{P}} + \frac{1}{m^2}\bigl(
+\frac{9}{8}O_{9}-O_7+\frac{1}{3}O_{6}+\frac{3}{2}O_{5}+\frac{7}{12}O_{4}+\frac{3}{4}O_{3}-\frac{1}{4}O_{2}-\frac{1}{2}O_{13}+\frac{3}{8}O_{12}+ \\ \nonumber 
&\qquad+\frac{3}{8}O_{11}+\frac{1}{8}O_{10}\bigr), \\
\tilde{O}_{35} &=- 2 O^{(0)}_{\bf{P}} + \frac{1}{m^2}\bigl(
-\frac{5}{8}O_{9}+O_7-\frac{1}{3}O_{6}-O_5-\frac{1}{12}O_4-\frac{3}{4}O_{3}+\frac{5}{4}O_{2}+O_{13}+\frac{1}{8}O_{12}+ \\ \nonumber 
&\qquad+\frac{1}{8}O_{11}-\frac{1}{8}O_{10}+\frac{1}{2}O_{1}\bigr), \\
\tilde{O}_{36} &= 6 O^{(0)}_{\bf{P}} + \frac{1}{m^2}\bigl(
-\frac{33}{8}O_{9}-3O_7-\frac{1}{2}O_{6}-\frac{3}{2}O_{5}-\frac{5}{4}O_{4}-\frac{9}{4}O_{3}+\frac{3}{4}O_{2}+\frac{3}{2}O_{13}-\frac{3}{8}O_{12}+ \\ \nonumber 
&\qquad-\frac{3}{8}O_{11}-\frac{9}{8}O_{10}-\frac{3}{2}O_{1}\bigr), \\
\tilde{O}_{37} &= - 6 O^{(0)}_{\bf{P}} + \frac{1}{m^2}\bigl(
+ \frac{9}{8}O_{9}+3O_7+\frac{3}{4}O_{4}+\frac{3}{4}O_{3}+\frac{3}{4}O_{2}+\frac{3}{8}O_{12}+\frac{3}{8}O_{11}-\frac{3}{8}O_{10}+\frac{3}{2}O_{1}\bigr), \\
\tilde{O}_{38} &=O^{(0)}_{\bf{P}} + \frac{1}{m^2}\bigl(
-\frac{1}{2}O_{9}+O_7-\frac{1}{12}O_6-\frac{1}{4}O_5+\frac{1}{24}O_{4}+\frac{1}{8}O_{3}+\frac{1}{8}O_{2}-\frac{1}{4}O_{10}+\frac{3}{8}O_{1}\bigr), \\
\tilde{O}_{39} &= -O^{(0)}_{\bf{P}} + \frac{1}{m^2}\bigl(-\frac{1}{4}O_{9}+O_8+3O_7-\frac{1}{12}O_6-\frac{1}{4}O_5+\frac{1}{24}O_{4}-\frac{1}{8}O_{3}-\frac{1}{8}O_{2}-\frac{1}{2}O_{13}+ \\ \nonumber 
&\qquad-\frac{1}{2}O_{12}-\frac{5}{4}O_{11}+\frac{1}{4}O_{10}-\frac{1}{8}O_{1}\bigr), \\
\tilde{O}_{40} &= -O^{(0)}_{\bf{P}} + \frac{1}{m^2}\bigl(
-\frac{11}{8}O_{9}-O_7-\frac{1}{3}O_{6}-O_5-\frac{7}{12}O_{4}-\frac{5}{4}O_{3}+O_2+\frac{5}{4}O_{13}-\frac{1}{8}O_{12}+ \\ \nonumber 
&\qquad-\frac{1}{8}O_{11}-\frac{3}{8}O_{10}-\frac{1}{2}O_{1}\bigr), \\
\tilde{O}_{41} &= -O^{(0)}_{\bf{P}} + \frac{1}{m^2}\bigl(
-\frac{17}{8}O_{9}-O_8-5O_7-\frac{1}{2}O_{6}-\frac{3}{2}O_{5}-\frac{3}{4}O_{4}-\frac{5}{2}O_{3}+\frac{9}{4}O_{2}+\frac{11}{4}O_{13}+ \\ \nonumber 
&\qquad+\frac{5}{8}O_{12}+\frac{11}{8}O_{11}-\frac{7}{8}O_{10}-\frac{1}{2}O_{1}\bigr), \\
\tilde{O}_{42} &= - O^{(0)}_{\bf{P}} + \frac{1}{m^2}\bigl(
+\frac{1}{8}O_{9}-O_7-\frac{1}{4}O_{4}-\frac{1}{4}O_{3}-\frac{1}{4}O_{2}-\frac{1}{8}O_{12}-\frac{1}{8}O_{11}+\frac{1}{8}O_{10}-\frac{1}{2}O_{1}\bigr), \\
\tilde{O}_{43} &= O^{(0)}_{\bf{P}} + \frac{1}{m^2}\bigl(
+\frac{11}{8}O_{9}+O_7+\frac{1}{3}O_{6}+O_5+\frac{7}{12}O_{4}+\frac{5}{4}O_{3}-\frac{3}{4}O_{2}-O_{13}+\frac{1}{8}O_{12}+ \\ \nonumber 
&\qquad+\frac{1}{8}O_{11}+\frac{3}{8}O_{10}+\frac{1}{2}O_{1}\bigr), \\
\tilde{O}_{44} &= O^{(0)}_{\bf{P}} + \frac{1}{m^2}\bigl(
+\frac{11}{8}O_{9}+O_7+\frac{1}{3}O_{6}+O_5+\frac{7}{12}O_{4}+\frac{5}{4}O_{3}-\frac{3}{4}O_{2}-O_{13}+\frac{1}{8}O_{12}+ \\ \nonumber 
&\qquad+\frac{1}{8}O_{11}+\frac{3}{8}O_{10}+\frac{1}{2}O_{1}\bigr), \\
    \tilde{O}_{45} &= O^{(0)}_{\bf{P}} + \frac{1}{m^2}\bigl(
    +\frac{11}{8}O_{9}+O_7+\frac{1}{3}O_{6}+O_5+\frac{7}{12}O_{4}+\frac{5}{4}O_{3}-\frac{3}{4}O_{2}-O_{13}+\frac{1}{8}O_{12}+ \\ \nonumber 
&\qquad+\frac{1}{8}O_{11}+\frac{3}{8}O_{10}+\frac{1}{2}O_{1}\bigr), \\
\tilde{O}_{46} &= -O^{(0)}_{\bf{P}} + \frac{1}{m^2}\bigl(
+\frac{1}{8}O_{9}-O_7-\frac{1}{4}O_{4}-\frac{1}{4}O_{3}-\frac{1}{4}O_{2}-\frac{1}{8}O_{12}-\frac{1}{8}O_{11}+\frac{1}{8}O_{10}-\frac{1}{2}O_{1}\bigr), \\
\tilde{O}_{47} &= O^{(0)}_{\bf{P}} + \frac{1}{m^2}\bigl( 
+\frac{5}{8}O_{9}-O_8-3O_7+\frac{1}{6}O_{6}+\frac{1}{2}O_{5}+\frac{5}{12}O_{4}+\frac{1}{2}O_{2}+\frac{1}{2}O_{13}+\frac{7}{8}O_{12}+ \\ \nonumber 
&\qquad+\frac{13}{8}O_{11}-\frac{1}{8}O_{10}+\frac{1}{2}O_{1}\bigr), \\
\tilde{O}_{48} &= O^{(0)}_{\bf{P}} + \frac{1}{m^2}\bigl( +\frac{17}{8}O_{9}+O_8+5O_7+\frac{1}{2}O_{6}+\frac{3}{2}O_{5}+\frac{3}{4}O_{4}+\frac{5}{2}O_{3}-2O_2-\frac{5}{2}O_{13} + \\ \nonumber 
&\qquad -\frac{5}{8}O_{12}-\frac{11}{8}O_{11}+\frac{7}{8}O_{10}+\frac{1}{2}O_{1}\bigr), \\
\tilde{O}_{49} &= O^{(0)}_{\bf{P}} + \frac{1}{m^2}\bigl(
+\frac{11}{8}O_{9}+O_7+\frac{1}{3}O_{6}+O_5+\frac{7}{12}O_{4}+\frac{5}{4}O_{3}-\frac{3}{4}O_{2}-O_{13}+\frac{1}{8}O_{12}+ + \\ \nonumber 
&\qquad\frac{1}{8}O_{11}+\frac{3}{8}O_{10}+\frac{1}{2}O_{1}\bigr), \\
\tilde{O}_{50} &= O^{(0)}_{\bf{P}} + \frac{1}{m^2}\bigl( +\frac{1}{4}O_{9}+O_8+3O_7-\frac{1}{12}O_6-\frac{1}{4}O_5+\frac{1}{24}O_{4}-\frac{1}{8}O_{3}-\frac{1}{8}O_{2}-\frac{1}{2}O_{13}+ \\ \nonumber 
&\qquad-\frac{1}{2}O_{12}-\frac{5}{4}O_{11}+\frac{1}{4}O_{10}-\frac{1}{8}O_{1}\bigr), \\
\tilde{O}_{51} &= - 3 O^{(0)}_{\bf{P}} + \frac{1}{m^2}\bigl(
-\frac{3}{2}O_{9}-O_7-\frac{1}{12}O_6-\frac{1}{4}O_5-\frac{5}{24}O_{4}-\frac{5}{8}O_{3}+\frac{1}{8}O_{2}+\frac{1}{4}O_{13}-\frac{1}{2}O_{10}-\frac{3}{8}O_{1}\bigr), \\
\tilde{O}_{52} &= 3 O^{(0)}_{\bf{P}} + \frac{1}{m^2}\bigl(
+O_7-\frac{1}{12}O_6-\frac{1}{4}O_5+\frac{1}{24}O_{4}+\frac{1}{8}O_{3}+\frac{1}{8}O_{2}-\frac{1}{4}O_{10}+\frac{3}{8}O_{1}\bigr), \\
\tilde{O}_{53} &= -3 O^{(0)}_{\bf{P}} + \frac{1}{m^2}\bigl(
-3O_9-3O_8-9O_7-\frac{5}{12}O_{6}-\frac{5}{4}O_{5}-\frac{19}{24}O_{4}-\frac{13}{8}O_{3}+\frac{21}{8}O_{2}+\frac{15}{4}O_{13}+ \\ \nonumber 
&\qquad+\frac{3}{2}O_{12}+\frac{15}{4}O_{11}-\frac{5}{4}O_{10}+\frac{3}{8}O_{1}\bigr), \\
\tilde{O}_{54} &= -2 O^{(0)}_{\bf{P}} + \frac{1}{m^2}\bigl(
+\frac{27}{8}O_{9}+O_7+\frac{5}{6}O_{6}+\frac{5}{2}O_{5}+\frac{13}{12}O_{4}+ \\ \nonumber 
&\qquad+\frac{11}{4}O_{3}-\frac{9}{4}O_{2}-\frac{5}{2}O_{13}+\frac{1}{8}O_{12}+\frac{1}{8}O_{11}+\frac{11}{8}O_{10}+\frac{1}{2}O_{1}\bigr), \\
\tilde{O}_{55} &=6 O^{(0)}_{\bf{P}} + \frac{1}{m^2}\bigl(
+\frac{15}{8}O_{9}-3O_7+\frac{1}{2}O_{6}+\frac{3}{2}O_{5}-\frac{1}{4}O_{4}+\frac{3}{4}O_{3}-\frac{9}{4}O_{2}-\frac{3}{2}O_{13}-\frac{3}{8}O_{12}+ \\ \nonumber 
&\qquad-\frac{3}{8}O_{11}-\frac{9}{8}O_{10}-\frac{3}{2}O_{1}\bigr), \\
\tilde{O}_{56} &=-6 O^{(0)}_{\bf{P}} + \frac{1}{m^2}\bigl(
+\frac{21}{8}O_{9}-3O_7+\frac{1}{2}O_{6}+\frac{3}{2}O_{5}+\frac{1}{2}O_{4}+\frac{3}{8}O_{12}+\frac{3}{8}O_{11}+\frac{9}{8}O_{10}-\frac{3}{4}O_{1}\bigr), \\
\tilde{O}_{57} &= 6 O^{(0)}_{\bf{P}} + \frac{1}{m^2}\bigl(
+\frac{27}{8}O_{9}+3O_7+\frac{1}{2}O_{6}+\frac{3}{2}O_{5}+\frac{1}{2}O_{4}+3O_3-3O_2-3O_{13}-\frac{3}{8}O_{12}+ \\ \nonumber 
&\qquad-\frac{3}{8}O_{11}+\frac{15}{8}O_{10}+\frac{3}{4}O_{1}\bigr), \\
\tilde{O}_{58} &= 2 O^{(0)}_{\bf{P}} + \frac{1}{m^2}\bigl(
-\frac{15}{8}O_{9}-O_7-\frac{1}{2}O_{6}-\frac{3}{2}O_{5}-\frac{1}{4}O_{4}-\frac{9}{4}O_{3}+\frac{9}{4}O_{2}+2O_{13}+\frac{3}{8}O_{12}+\frac{3}{8}O_{11}-\frac{7}{8}O_{10}\bigr), \\
\tilde{O}_{59} &=-6 O^{(0)}_{\bf{P}} + \frac{1}{m^2}\bigl(
-\frac{21}{8}O_{9}+2O_8+7O_7-\frac{1}{3}O_{6}-2O_5-\frac{5}{6}O_{4}-\frac{1}{2}O_{3}-O_2-O_{13} + \\ \nonumber 
&\qquad-\frac{15}{8}O_{12}-\frac{27}{8}O_{11}+\frac{7}{8}O_{10}-\frac{3}{4}O_{1}\bigr), \\
\tilde{O}_{60} &= 6 O^{(0)}_{\bf{P}}  + \frac{1}{m^2}\bigl(
-\frac{3}{8}O_{9}+3O_7+\frac{3}{4}O_{4}+\frac{3}{4}O_{3}+\frac{3}{4}O_{2}+\frac{3}{8}O_{12}+\frac{3}{8}O_{11}-\frac{3}{8}O_{10}+\frac{3}{2}O_{1}\bigr), \\
\tilde{O}_{61} &= -6 O^{(0)}_{\bf{P}} + \frac{1}{m^2}\bigl(
-\frac{39}{8}O_{9}-2O_8-7O_7-\frac{7}{6}O_6-\frac{5}{2}O_{5}-\frac{13}{6}O_{4}-\frac{5}{2}O_{3}+\frac{5}{2}O_{2}+4O_{13}+ \\ \nonumber 
&\qquad+\frac{3}{8}O_{12}+\frac{15}{8}O_{11}-\frac{19}{8}O_{10}-\frac{3}{4}O_{1}\bigr), \\
\tilde{O}_{62} &= 2 O^{(0)}_{\bf{P}}+ + \frac{1}{m^2}\bigl(
\frac{1}{8}O_{9}+2O_8+5O_7-\frac{1}{6}O_{5}-\frac{1}{6}O_{3}-\frac{1}{2}O_{2}-O_{13}-\frac{9}{8}O_{12}-\frac{21}{8}O_{11}+ \\ \nonumber 
&\qquad+\frac{1}{8}O_{10}-\frac{3}{4}O_{1}\bigr), \\
\tilde{O}_{63} &=- 6 O^{(0)}_{\bf{P}} + \frac{1}{m^2}\bigl(
-\frac{9}{8}O_{9}+O_7-\frac{1}{3}O_{6}-\frac{3}{2}O_{5}-\frac{7}{12}O_{4}-\frac{3}{4}O_{3}+\frac{1}{4}O_{2}+\frac{1}{2}O_{13}-\frac{3}{8}O_{12}+ \\ \nonumber 
&\qquad-\frac{3}{8}O_{11}-\frac{1}{8}O_{10}\bigr), \\
\tilde{O}_{64} &= 6 O^{(0)}_{\bf{P}} + \frac{1}{m^2}\bigl(
-\frac{3}{8}O_{9}+O_7+\frac{1}{2}O_{5}+\frac{1}{2}O_{4}+\frac{1}{2}O_{3}+\frac{1}{2}O_{2}+\frac{3}{8}O_{12}+\frac{3}{8}O_{11}+\frac{1}{8}O_{10}+\frac{3}{4}O_{1}\bigr), \\
\tilde{O}_{65} &= -6 O^{(0)}_{\bf{P}} + \frac{1}{m^2}\bigl(
-\frac{39}{8}O_{9}-6O_8-17O_7-O_6-\frac{5}{2}O_{5}-\frac{5}{4}O_{3}-\frac{13}{4}O_{3}+\frac{19}{4}O_{2}+\frac{13}{2}O_{13}+ \\ \nonumber 
&\qquad+\frac{27}{8}O_{12}+\frac{63}{8}O_{11}-\frac{19}{8}O_{10}+\frac{3}{2}O_{1}\bigr), \\
\tilde{O}_{66} &= -2 O^{(0)}_{\bf{P}} + \frac{1}{m^2}\bigl(
-\frac{1}{2}O_{9}-2O_8-6O_7+\frac{1}{6}O_{6}+\frac{1}{3}O_{5}-\frac{1}{12}O_4+\frac{1}{12}O_{3}+\frac{1}{4}O_{2}+O_{13}+ \\ \nonumber 
&\qquad+O_{12}+\frac{5}{2}O_{11}-\frac{1}{2}O_{10}+\frac{1}{4}O_1\bigr), \\
\tilde{O}_{67} &= 6 O^{(0)}_{\bf{P}} + \frac{1}{m^2}\bigl(
+3O_9+2O_7+\frac{1}{6}O_{6}+O_5+\frac{5}{12}O_{4}+\frac{7}{4}O_{3}-\frac{1}{4}O_{2}-\frac{1}{2}O_{13}+O_{10}+\frac{3}{4}O_{1}\bigr), \\
\tilde{O}_{68} &= -6 O^{(0)}_{\bf{P}} + \frac{1}{m^2}\bigl(
-2O_7+\frac{1}{2}O_{5}-\frac{1}{4}O_{4}-\frac{1}{4}O_{3}-\frac{1}{4}O_{2}+\frac{1}{2}O_{10}-\frac{3}{4}O_{1}\bigr), \\
\tilde{O}_{69} &= 6 O^{(0)}_{\bf{P}} + \frac{1}{m^2}\bigl(
+6O_9+6O_8+18O_7+O_6+\frac{5}{2}O_{5}+\frac{7}{4}O_{4}+\frac{13}{4}O_{3}-\frac{21}{4}O_{2}-\frac{15}{2}O_{13}+ \\ \nonumber 
&\qquad-3O_{12}-\frac{15}{2}O_{11}+\frac{5}{2}O_{10}-\frac{3}{4}O_{1}\bigr), \\
\tilde{O}_{70} &= 2 O^{(0)}_{\bf{P}} + \frac{1}{m^2}\bigl(
-\frac{15}{8}O_{9}-O_7-\frac{1}{2}O_{6}-\frac{3}{2}O_{5}-\frac{1}{4}O_{4}-\frac{9}{4}O_{3}+\frac{9}{4}O_{2}+2O_{13}+\frac{3}{8}O_{12}+ \\ \nonumber 
&\qquad+\frac{3}{8}O_{11}-\frac{7}{8}O_{10}\bigr), \\
\tilde{O}_{71} &= - 6 O^{(0)}_{\bf{P}} + \frac{1}{m^2}\bigl(
-\frac{45}{8}O_{9}-4O_8-9O_7-O_6-4O_5-2O_4-\frac{5}{2}O_{3}+4O_2+6O_{13}+\frac{9}{8}O_{12}+ \\ \nonumber 
&\qquad+\frac{33}{8}O_{11}-\frac{5}{8}O_{8}+\frac{3}{4}O_{1}\bigr), \\
\tilde{O}_{72} &= 6 O^{(0)}_{\bf{P}} + \frac{1}{m^2}\bigl(
-\frac{3}{8}O_{9}+3O_7+\frac{3}{4}O_{4}+\frac{3}{4}O_{3}+\frac{3}{4}O_{2}+\frac{3}{8}O_{12}+\frac{3}{8}O_{11}-\frac{3}{8}O_{10}+\frac{3}{2}O_{1}\bigr), \\
\tilde{O}_{73} &= -6 O^{(0)}_{\bf{P}} + \frac{1}{m^2}\bigl( -\frac{15}{8}O_{9}+4O_8+9O_7-\frac{1}{2}O_{6}-\frac{1}{2}O_{5}-O_4-\frac{1}{2}O_{3}-\frac{5}{2}O_{2}-3O_{13}+ \\ \nonumber 
&\qquad-\frac{21}{8}O_{12}-\frac{45}{8}O_{11}-\frac{7}{8}O_{10}-\frac{9}{4}O_{1}\bigr), \\
\tilde{O}_{74} &= 2O^{(0)}_{\bf{P}} + \frac{1}{m^2}\bigl(
+\frac{1}{8}O_{9}+2O_8+5O_7-\frac{1}{6}O_{5}-\frac{1}{6}O_{3}-\frac{1}{2}O_{2}-O_{13}-\frac{9}{8}O_{12}-\frac{21}{8}O_{11}+ \\ \nonumber 
&\qquad+\frac{1}{8}O_{10}-\frac{3}{4}O_{1}\bigr), \\
\tilde{O}_{75} &=- 6 O^{(0)}_{\bf{P}} + \frac{1}{m^2}\bigl(
-\frac{33}{8}O_{9}-6O_8-15O_7-O_6-\frac{7}{2}O_{5}-\frac{7}{4}O_{4}-\frac{11}{4}O_{3}+\frac{21}{4}O_{2}+\frac{15}{2}O_{13}+ \\ \nonumber 
&\qquad+\frac{21}{8}O_{12}+\frac{57}{8}O_{11}-\frac{13}{8}O_{10}+\frac{3}{2}O_{1}\bigr), \\
\tilde{O}_{76} &= 6 O^{(0)}_{\bf{P}} + \frac{1}{m^2}\bigl(
-\frac{3}{8}O_{9}+O_7+\frac{1}{2}O_{5}+\frac{1}{2}O_{4}+\frac{1}{2}O_{3}+\frac{1}{2}O_{2}+\frac{3}{8}O_{12}+\frac{3}{8}O_{11}+\frac{1}{8}O_{10}+\frac{3}{4}O_{1}\bigr), \\
\tilde{O}_{77} &= - 6 O^{(0)}_{\bf{P}} + \frac{1}{m^2}\bigl(
-\frac{15}{8}O_{9}-O_7-\frac{1}{3}O_{6}-\frac{1}{2}O_{5}-\frac{1}{12}O_4-\frac{5}{4}O_{3}-\frac{1}{4}O_{2}-\frac{1}{2}O_{13}+\frac{3}{8}O_{12}+ \\ \nonumber 
&\qquad+\frac{3}{8}O_{11}-\frac{7}{8}O_{10}\bigr), \\
\tilde{O}_{78} &= -2 O^{(0)}_{\bf{P}} + \frac{1}{m^2}\bigl(
-\frac{1}{2}O_{9}-2O_8-6O_7+\frac{1}{6}O_{6}+\frac{1}{3}O_{5}-\frac{1}{12}O_4+\frac{1}{12}O_{3}+\frac{1}{4}O_{2}+O_{13}+ \\ \nonumber 
&\qquad+O_{12}+\frac{5}{2}O_{11}-\frac{1}{2}O_{10}+\frac{1}{4}O_1 \bigr), \\
\tilde{O}_{79} &= 6 O^{(0)}_{\bf{P}} + \frac{1}{m^2}\bigl(
+6O_9+6O_8+18O_7+\frac{5}{6}O_{6}+3O_5+\frac{19}{12}O_{4}+\frac{15}{4}O_{3}-\frac{21}{4}O_{2}-\frac{15}{2}O_{13}+ \\ \nonumber 
&\qquad -3O_{12}-\frac{15}{2}O_{11}+\frac{5}{2}O_{10}-\frac{3}{4}O_{1}\bigr), \\
\tilde{O}_{80} &= - 6 O^{(0)}_{\bf{P}} + \frac{1}{m^2}\bigl(
-2O_7+\frac{1}{2}O_{5}-\frac{1}{4}O_{4}-\frac{1}{4}O_{3}-\frac{1}{4}O_{2}+\frac{1}{2}O_{10}-\frac{3}{4}O_{1}\bigr), \\
\tilde{O}_{81} &= 6 O^{(0)}_{\bf{P}} + \frac{1}{m^2}\bigl(
+3O_9+2O_7+\frac{1}{3}O_{6}+\frac{1}{2}O_{5}+\frac{7}{12}O_{4}+\frac{5}{4}O_{3}-\frac{1}{4}O_{2}-\frac{1}{2}O_{13}+O_{10}+\frac{3}{4}O_{1}\bigr), \\
\tilde{O}_{82} &= -2 O^{(0)}_{\bf{P}} + \frac{1}{m^2}\bigl(
-2O_8-4O_7+\frac{1}{2}O_{3}+\frac{1}{2}O_{2}+O_{13}+O_{12}+\frac{5}{2}O_{11}+O_1\bigr), \\
\tilde{O}_{83} &= 6 O^{(0)}_{\bf{P}} + \frac{1}{m^2}\bigl(
-2O_7+\frac{1}{6}O_{6}+\frac{1}{2}O_{5}-\frac{1}{12}O_4-\frac{1}{4}O_{3}+\frac{1}{4}O_{2}+\frac{1}{2}O_{13}-\frac{3}{4}O_{1} \bigr), \\
\tilde{O}_{84} &= - 6 O^{(0)}_{\bf{P}} + \frac{1}{m^2}\bigl(
+\frac{3}{2}O_{9} \bigr), \\
\tilde{O}_{85} &= 6 O^{(0)}_{\bf{P}} + \frac{1}{m^2}\bigl(
+3O_9+6O_8+14O_7+\frac{5}{6}O_{6}+\frac{5}{2}O_{5}+\frac{13}{12}O_{4}+\frac{7}{4}O_{3}-\frac{19}{4}O_{2}-\frac{13}{2}O_{13}+ \\ \nonumber 
&\qquad-3O_{12}-\frac{15}{2}O_{11}+\frac{3}{2}O_{10}-\frac{9}{4}O_{1} \bigr), \\
\tilde{O}_{86} &= - 12 O^{(0)}_{\bf{P}}+ \frac{1}{m^2}\bigl( -\frac{9}{4}O_{9}-2O_8-6O_7+\frac{1}{3}O_{6}-O_5-\frac{1}{6}O_{4}-O_3+2O_2+3O_{13}+\frac{3}{4}O_{12}+ \\ \nonumber 
&\qquad+\frac{9}{4}O_{11}+\frac{1}{4}O_{10} \bigr), \\
\tilde{O}_{87} &= -12 O^{(0)}_{\bf{P}} + \frac{1}{m^2}\bigl( 
+\frac{3}{4}O_{9}+\frac{1}{2}O_{6}+\frac{3}{2}O_{5}+\frac{5}{4}O_{4}+\frac{3}{4}O_{3}+\frac{3}{4}O_{2}+\frac{3}{4}O_{12}+\frac{3}{4}O_{11}+\frac{3}{4}O_{10}+\frac{3}{4}O_{1} \bigr), \\
\tilde{O}_{88} &= -12 O^{(0)}_{\bf{P}} + \frac{1}{m^2}\bigl(
+6O_9+4O_7-\frac{1}{3}O_{6}-O_5+\frac{1}{6}O_{4}+\frac{1}{2}O_{3}+\frac{3}{2}O_{2}+O_{13}-O_{10}+\frac{3}{2}O_{1} \bigr), \\
\tilde{O}_{89} &= - 12 O^{(0)}_{\bf{P}} + \frac{1}{m^2}\bigl(
+\frac{9}{8}O_{9}+O_7+\frac{1}{6}O_{6}+\frac{1}{2}O_{5}+\frac{2}{3}O_{4}+\frac{1}{2}O_{3}-\frac{1}{2}O_{13}+\frac{3}{8}O_{12}+\frac{3}{8}O_{11}+ \\ \nonumber 
&\qquad+\frac{1}{8}O_{10}+\frac{3}{4}O_{1} \bigr), \\
\tilde{O}_{90} &= 24 O^{(0)}_{\bf{P}} + \frac{1}{m^2}\bigl(
-\frac{9}{2}O_{9}-4O_8-12O_7-2O_5-O_4-2O_3+4O_2+6O_{13}+\frac{3}{2}O_{12}+\frac{9}{2}O_{11}+\frac{1}{2}O_{10} \bigr), \\
\tilde{O}_{91} &= 24 O^{(0)}_{\bf{P}} + \frac{1}{m^2}\bigl(
-3O_9-2O_8-6O_7-\frac{1}{6}O_{6}-\frac{5}{2}O_{5}-\frac{17}{12}O_{4}-\frac{7}{4}O_{3}+\frac{5}{4}O_{2}+3O_{13}+ \\ \nonumber 
&\qquad +\frac{3}{2}O_{11}-\frac{1}{2}O_{10}-\frac{3}{4}O_{1} \bigr), \\
\tilde{O}_{92} &= 12 O^{(0)}_{\bf{P}} + \frac{1}{m^2}\bigl(
-\frac{33}{4}O_{9}-6O_7-\frac{3}{2}O_{4}-\frac{3}{2}O_{3}-\frac{3}{2}O_{2}-\frac{3}{4}O_{12}-\frac{3}{4}O_{11}+\frac{3}{4}O_{10}-3O_1 \bigr). \label{eq:98}
\end{align}}
\section{Details of the calculation of the boost correction from Poincaré Algebra}
\label{appendix:boost}
In this appendix, we present an outline of the calculation process leading to the result of Eq. (\ref{delta_V_P_poincarè}) for the 3N boost correction. Further details may be found in Ref. \cite{nasoni}. 
Starting from Eq. (\ref{boost correction}), we calculate the matrix elements of $\delta V(\bm{P})$ between 3N initial and final states $\Psi$ and $\Psi'$, i.e. $\bra{\Psi} \delta V(\bm{P}) \ket{\Psi'}$.

First of all, we remind that the leading contact potential $V^{(0)}$ is an identity of spin and isospin, thus its matrix elements between 3N eigenstates $\ket{\bm{p_1}, \sigma_1, \tau_1; \bm{p_2}, \sigma_2, \tau_2; \bm{p_3}, \sigma_3, \tau_3}$ (see Appendix \ref{appendix:orthoclosure})  are given by
\begin{eqnarray}
	    \bra{\bm{p_1}, \sigma_1, \tau_1; \bm{p_2}, \sigma_2, \tau_2; \bm{p_3}, \sigma_3, \tau_3} &V^{(0)} &\ket{\bm{p'_1}, \sigma'_1, \tau'_1; \bm{p'_2}, \sigma'_2, \tau'_2; \bm{p'_3}, \sigma'_3, \tau'_3} = \nonumber\\
     & = & E_0 \delta(\bm{P} - \bm{P'}) \prod_{\nu=1}^{3} \delta_{\sigma_{\nu} \sigma_{\nu}'}  \delta_{\tau_{\nu} \tau_{\nu}'} + \text{exchange terms} ;   \label{contact U0}  
\end{eqnarray}

We will use Jacobi coordinates defined in Eq. (\ref{Jacobi constituents}).

We may adopt the following notation conventions: operators of momentum and position are indicated by $\bm{\pi_{a,b}}$, $\bm{\rho_{a,b}}$, respectively, and their eigenvalues with $\bm{q_{a,b}}$, $\bm{\xi_{a,b}}$, respectively ( i.e., $\bm{\pi_{a,b}}\ket{\bm{q_{a,b}}}=\bm{q_{a,b}}\ket{\bm{q_{a,b}}}$ and 
 $\bm{\rho_{a,b}} \ket{\bm{\xi_{a,b}}} = \bm{\xi_{a,b}} \ket{\bm{\xi_{a,b}}}$). We introduce the shorthands  $\bm{q} \equiv (\bm{q_a}, \bm{q_b}$) and $\bm{q'} \equiv (\bm{q_a'}, \bm{q_b'})$, $\bm{dq} \equiv \bm{dq_a}\bm{dq_b}$ and $\bm{dq'} \equiv \bm{dq_a'}\bm{dq_b'}$, $\Psi \equiv \Psi(\bm{P}, \bm{q})$ and $\Psi' \equiv \Psi' (\bm{P'}, \bm{q'})$. The notation $\bigl[(\bm{q'}, \bm{q''}) \rightarrow (\bm{q''}, \bm{q})\bigr]$  indicates substitution of $\bm{q'}$ with $\bm{q''}$ and $\bm{q''}$ with $\bm{q}$ that precedes it inside the round brackets. We consider the particles to be identical in the initial and final states $\Psi$ and $\Psi'$, while we treat them as they were distinguishable in the intermediate states (see Appendix \ref{appendix:orthoclosure}). 

During calculations we can ignore permutation terms, using orthogonality and closure relations given by Eqs. (\ref{closurejacobi:P,p})-(\ref{orthojacobi:R,xi}).

Therefore, we may rewrite Eq. (\ref{contact U0}) using Jacobi coordinates as
{\small \begin{equation}
    \label{U_0 braket}
	\bra{\bm{P}, \bm{q_a}, \bm{q_b}} V^{(0)}\ket{\bm{P'}, \bm{q'_a}, \bm{q'_b}} = E_0 \delta(\bm{P} - \bm{P'}),
\end{equation}}
suppressing spin and isospin indices in $3N$ states labels and neglecting the spin-isospin Kronecker deltas $\delta_{\sigma \sigma'} \delta_{\tau \tau'}$.

The factor $E_0$ can be safely ignored during the calculation process and later reintroduced, as it merely acts as a global factor. 

To proceed with the calculation of $ \bra{\Psi'} \delta V(\bm P) \ket{\Psi} $ , we write $\chizero  \equiv {{\chi}} ^{(1)} + {{\chi}} ^{(2)} + {{\chi}} ^{(3)}$, with
{\small \begin{align}
	\label{chizeroone jacobi}
	\chizeroone &= -\frac{1}{4 (3m)^2} \biggl[ \bm{\rho_a} \cdot \bm{P} \bm{\pi_a} \cdot \bm{P} + \bm{\rho_b} \cdot \bm{P} \bm{\pi_b} \cdot \bm{P}\biggr] + H. c.,  \\
	\label{chizerotwo jacobi}
	\chizerotwo &=  \frac{1}{12 m^2} \biggl[ \bm{\rho_a} \cdot \bm{P} \bm{\pi_a} \cdot \bm{\pi_b} - \frac{1}{2}\bm{\rho_b} \cdot \bm{P} {\bm{\pi_b}}^2 + \frac{2}{3} \bm{\rho_b} \cdot \bm{P} {\bm{\pi_a}}^2\biggr] + H. c.,  \\
	\label{chizerothree jacobi}
 	\chizerothree &= - \frac{1}{6 m^2} \Biggl[\bm{s_a} \wedge \bm{P} \cdot \bm{\pi_a} + \bm{s_b} \wedge \bm{P} \cdot \bm{\pi_b} \Biggr],
\end{align}}
where $\bm{s_a} \equiv \bm{s_1} - \bm{s_2}$ and $\bm{s_b} \equiv \bm{s_3} - \frac{\bm{s_1} + \bm{s_2}}{2}$. 

We find the following intermediate results,
{\small	\begin{align}
		\label{boostcorr P^2}
		\bra{\Psi '} - \frac{\bm{P}^2 V^{(0)}}{2 (3m)^2} \ket{\Psi} &= - \frac{E_0}{2(3 m^2)} \integralbtwtwostates \bm{P}^2 \\
		\label{boostcorr chizeroone}
		-i	\bra{\Psi '} [\chizeroone, V^{(0)}] \ket{\Psi }&=- \frac{E_0}{4(3m)^2}  \integralbtwtwostates 4\bm{P}^2 \\
		\label{boostcorr chizerotwo}
		- i \bra{\Psi '}[\chizerotwo, V^{(0)}] \ket{\Psi} &= 0 \\
		\label{boostcorr chizerothree}
		- i \bra{\Psi '} [\chizerothree, V^{(0)}] \ket{\Psi} &= -i \frac{E_0}{6 m^2} \integralbtwtwostates \bigl[\bm{P} \wedge (\bm{q}_a - \bm{q}_a') \cdot \bm{s}_a + \bm{P} \wedge (\bm{q_b} - \bm{q_b'}) \cdot \bm{s}_b \bigr],
	\end{align}}
leading to the final result in Jacobi coordinates,
{\small
\begin{equation}
	\label{boost jacobi}
		\bra{\Psi '} \delta V({\bm P}) \ket{\Psi} = -\frac{E_0}{6m^2} \integralbtwtwostates \biggl[\bm{P^2} + i \bm{P} \wedge (\bm{q_a} - \bm{q_a'}) \cdot \bm{s_a} + i \bm{P} \wedge (\bm{q_b} - \bm{q_b'}) \cdot \bm{s_b} \biggr]. 
\end{equation}}

In matrix elements such as those appearing in the above Eqs. (\ref{boostcorr P^2}), (\ref{boostcorr chizeroone}), (\ref{boostcorr chizerotwo}), (\ref{boostcorr chizerothree}), exchange terms will always be present, analogously as in Eqs. (\ref{states}), (\ref{ortho}), (\ref{closure}), due to the symmetry of the operators under the exchange of particles (see Appendix \ref{appendix:orthoclosure}). Such exchange terms will be implicitly understood during calculations.

Eq. (\ref{boostcorr P^2}) follows straightly from (\ref{U_0 braket}). 

In this appendix, we undertake the calculations leading to Eq. (\ref{boostcorr chizeroone}); Eq. (\ref{boostcorr chizerotwo}) and Eq. (\ref{boostcorr chizerothree}) are obtained with a similar procedure. 

It follows from Eq. (\ref{U_0 braket}) that, for any operator $\bm{X}$,  
{\small \begin{equation}
	\label{XU commutator btw psi'psi}
	\begin{split}
		\matrixelementpsiapexpsi{[\bm{X}, V^{(0)}]} = & \int \bm{dP} \bm{dP'} \bm{dq}\bm{dq'}\bm{dq''}  \Psi'^{*} \Psi \biggl[\bra{\bm{P'}, \bm{q'}} \bm{X} \ket{\bm {P}, \bm{q''}} -  \bra{\bm{P'}, \bm{q''}} \bm{X} \ket{\bm{P}, \bm{q}} \biggr]\\
		= & \int \bm{dP} \bm{dP'} \bm{dq}\bm{dq'}\bm{dq''}  \Psi'^{*} \Psi \biggl[\bra{\bm{P'}, \bm{q'}} \bm{X} \ket{\bm {P}, \bm{q''}} -  \bigl[ (\bm{q'}, \bm{q''}) \rightarrow (\bm{q''}, \bm{q})\bigr] \biggr].
	\end{split}
\end{equation}}

In Eq. (\ref{XU commutator btw psi'psi}) we identify  $\bm{X} = \chizeroone$, with  $\chizeroone = \bigl( \bm{\rho_a} \cdot \bm{P} \bm{\pi_a} \cdot \bm{P} +  \bm{\pi}_a \cdot \bm{P} \bm{\rho_a} \cdot \bm{P} \bigr) + \bigl(\bm{\rho_b} \cdot \bm{P} \bm{\pi_b} \cdot \bm{P} +  \bm{\pi}_b \cdot \bm{P} \bm{\rho_b} \cdot \bm{P} \bigr) $. Developing the calculation only for the component involving Jacobi variables $\bm{\rho_a}$, $\bm{\pi_a}$, we get
\begin{eqnarray}
    -i\bra{\Psi'} \biggl[-\frac{1}{4 (3m)^2} &\bigl(& \bm{\rho_a} \cdot \bm{P} \bm{\pi_a} \cdot \bm{P} +  \bm{\pi}_a \cdot \bm{P} \bm{\rho_a} \cdot \bm{P} \bigr)  , V^{(0)}\biggr] \ket{\Psi}= \nonumber\\
	 &= &  i \frac{1}{4(3m)^2} \int \bm{dP dP' dq dq' dq''}  \Psi'^{*} \Psi \biggl[\bra{\bm{P', q'}} \bigl[ \bm{\rho_a} \cdot \bm{P} \bm{\pi_a} \cdot \bm{P} +  \bm{\pi}_a \cdot \bm{P} \bm{\rho_a} \cdot \bm{P} \bigr] \ket{\bm{P, q''}} \nonumber\\
 & -&\bigl[ (\bm{q', q''}) \rightarrow (\bm{q'', q})\bigr] \biggr].	\label{chizeroone calc: first step}
\end{eqnarray}
We insert the coordinates closure relation (\ref{closurejacobi:R,xi}),
{\small \begin{equation}
	\label{chizeroone calc: coordinate closure step}
	\begin{split}
		&\bra{\bm{P', q'}} \bigl( \bm{\rho_a} \cdot \bm{P} \bm{\pi_a} \cdot \bm{P} +  \bm{\pi_a} \cdot \bm{P} \bm{\rho_a} \cdot \bm{P} \bigr) \ket{\bm{P, q''}} = \\
		&= \bra{\bm{P', q'}} \bm{\xi_a}  \ket{\bm{P, q''}} \cdot \bm{P}   (\bm{q''_a}  + \bm{q'_a}) \cdot \bm{P}   \\
		&= \bra{\bm{P', q'}} \mathbb{1}_{(\bm{R, \xi})} \bm{\xi_a}  \ket{\bm{P, q''}} \cdot \bm{P}   (\bm{q''_a}  + \bm{q'_a}) \cdot \bm{P}   \\
		&= \int \bm{d\xi_a}  \orthodelta{P}{P'} \orthodelta{q_b''}{q_b'} \expcoordmomvec{\xi_a}{q''_a}{q'_a} \bm{\xi}_a  \cdot \bm{P} (\bm{q''_a} + \bm{q'_a}) \cdot \bm{P},
	\end{split}
\end{equation}}
and we remind that since variables $\bm{\xi_a}$ and $\bm{q_a'}$ are canonically conjugate, it  holds
{ \small \begin{equation}
	\label{coordinate in momenta representation}
	\begin{split}
		\expcoordmomvec{\xi_a}{q''_a}{q'_a}  \bm{\xi_a} &= i \partialvecright{q_a'} \biggl[ \expcoordmomvec{\xi_a}{q_a''}{q_a'} \biggr], \\
		\expcoordmomvec{\xi_a}{q_a}{q_a''}  \bm{\xi_a} &= - i \partialvecright{q_a} \biggl[ \expcoordmomvec{\xi_a}{q_a}{q_a''} \biggr];
	\end{split}
\end{equation}}
therefore, by substituting Eq. (\ref{coordinate in momenta representation}) in Eq. (\ref{chizeroone calc: coordinate closure step}), and then Eq. (\ref{chizeroone calc: coordinate closure step}) in Eq. (\ref{chizeroone calc: first step}), and performing the integration of Eq. (\ref{chizeroone calc: first step}) with respect to $\bm{P'}$ and $\bm{q^{"}_{b}}$, we obtain
\begin{align}
		-i\bra{\Psi'} \biggl[-\frac{1}{4 (3m)^2}& \bigl( \bm{\rho_a} \cdot \bm{P} \bm{\pi_a} \cdot \bm{P} +  \bm{\pi_a} \cdot \bm{P} \bm{\rho_a} \cdot \bm{P} \bigr)  , V^{(0)} \biggr] \ket{\Psi} = \nonumber\\
		& = - \frac{1}{4(3m)^2} \int \bm{dP dq dq' dq_a'' d\xi_a}  \Psi'^{*} \Psi \Biggl\{ \partialvecright{q_a'} \biggl[ \expcoordmomvec{\xi_a}{q_a''}{q_a'} \biggr] \cdot \bm{P} (\bm{q''_a} + \bm{q'_a}) \cdot \bm{P} + \nonumber\\
		& \quad +  \bigl[ (\bm{q', q''}) \rightarrow (\bm{q'', q})\bigr] \Biggr\}.   \label{chizeroone calc: from coord to der}
\end{align}
We now integrate the first term in Eq. (\ref{chizeroone calc: from coord to der}) by parts with respect to $\bm{q_a'}$. The other term obtained by exchanging $\bigl[(\bm{q', q''}) \rightarrow (\bm{q'', q})\bigr]$ is treated in a similar way, integrating it by parts with respect to $\bm{q_a}$, and contributes in the same way. We obtain
{ \small \begin{equation}
	\label{chizeroone calc: integrating by part}
	\begin{split}
		& \int \bm{dP dq dq' dq_a'' d\xi_a} \Psi'^{*} \Psi  \partialvecright{q_a'} \biggl[ \expcoordmomvec{\xi_a}{q_a''}{q_a'} \biggl] \cdot \bm{P} (\bm{q''_a} + \bm{q'_a}) \cdot \bm{P} = \\
		&= \int \bm{dP dq dq' dq_a'' d\xi_a}  \bm{P} \cdot \partialvecright{q_a'}  \biggl[ \Psi'^{*} \Psi   \expcoordmomvec{\xi_a}{q_a''}{q_a'} (\bm{q''_a} + \bm{q'_a}) \cdot \bm{P} \biggr] \\
		& - \int \bm{dP dq dq' dq_a'' d\xi_a} \bm{P} \cdot \partialvecright{q_a'}  \biggl[ \Psi'^{*} \Psi    (\bm{q''_a} + \bm{q'_a}) \cdot \bm{P} \biggr] \expcoordmomvec{\xi_a}{q_a''}{q_a'}   \\
		& \equiv I_1 + I_2 =\int \bm{dP dq dq'} \Psi'^{*} \Psi \bm{P}^2,
	\end{split}
\end{equation}}
where
{\small \begin{align}
	\begin{split}
		I_1 &= \int \bm{dq_a'' d\xi_a}\quad \bm{P} \cdot \partialvecright{q_a'}  \biggl[\Psi'^{*} \Psi   \expcoordmomvec{\xi_a}{q_a''}{q_a'}  (\bm{q''_a} + \bm{q'_a}) \cdot \bm{P} \biggr] \\
		& \quad =  \bm{P} \cdot \partialvecright{q_a'}  \biggl[   \Psi'^{*} \Psi 2 \bm{q'_a} \cdot \bm{P} \biggr]; 
	\end{split}\\
	\begin{split}
		I_2 & = - \int \bm{dq_a'' d\xi_a}  \bm{P} \cdot \partialvecright{q_a'}  \biggl[ \Psi'^{*} \Psi    (\bm{q''_a} + \bm{q'_a}) \cdot \bm{P} \biggr] \expcoordmomvec{\xi_a}{q_a''}{q_a'}  \\
		&=  - I_1 + \Psi'^{*} \Psi \bm{P}^2.
	\end{split}
\end{align}}
We take into account both terms in Eq. (\ref{chizeroone calc: from coord to der}) and find
{\small \begin{equation}
		 - i\bra{\Psi'} \biggl[- \frac{1}{4 (3m)^2} \bigl( \bm{\rho_a} \cdot \bm{P} \bm{\pi}_a \cdot \bm{P} +  \bm{\pi}_a \cdot \bm{P} \bm{\rho_a} \cdot \bm{P} \bigr)  , V^{(0)} \biggr] \ket{\Psi}= \\
  = - \frac{1}{4(3m)^2} \int \bm{dP dq dq' } \Psi'^{*} \Psi 2 \bm{P}^2. 
\end{equation}}
The component in $\chi_0^{(1)}$ involving Jacobi variables $\bm{\rho_b}$, $\bm{\pi_b}$ makes an equal contribution. Thus, the final result is given by Eq. (\ref{boostcorr chizeroone}).\\
We observe that it would have been easier to write, instead of (\ref{coordinate in momenta representation}), 
\begin{equation}
\label{easier boost calc}
	\begin{split}
		\expcoordmomvec{\xi_a}{q''_a}{q'_a}  \bm{\xi_a} &= - i \partialvecright{q_a''} \biggl[ \expcoordmomvec{\xi_a}{q_a''}{q_a'} \biggr], \\
		\expcoordmomvec{\xi_a}{q_a}{q_a''}  \bm{\xi_a} &= + i \partialvecright{q_a''} \biggl[ 	\expcoordmomvec{\xi_a}{q_a}{q_a''} \biggr].
	\end{split}
\end{equation}
In this case, by substituting Eq. (\ref{easier boost calc}) in Eq. (\ref{chizeroone calc: coordinate closure step}), then Eq. (\ref{chizeroone calc: coordinate closure step}) in Eq. (\ref{chizeroone calc: first step}), and performing the integration by parts with respect to $\bm{q_b''}$, we obtain
{ \small \begin{equation}
	\begin{split}
		&-i\bra{\Psi'} \biggl[-\frac{1}{4 (3m)^2} \bigl( \bm{\rho_a} \cdot \bm{P} \bm{\pi_a} \cdot \bm{P} +  \bm{\pi_a} \cdot \bm{P} \bm{\rho_a} \cdot \bm{P} \bigr)  , V^{(0)} \biggr] \ket{\Psi} = \\
		& = - \frac{1}{4(3m)^2} \int \bm{dP dq dq' dq_a'' d\xi_a}  \Psi'^{*} \Psi \biggl[ \partialvecright{q_a''} \bigl( \expcoordmomvec{\xi_a}{q_a''}{q_a'} \bigr) \cdot \bm{P} (\bm{q''_a} + \bm{q'_a}) \cdot \bm{P} \\
		& \quad +  \bigl[ (\bm{q', q''}) \rightarrow (\bm{q'', q})\bigr] \biggr] \\
		& = - \frac{1}{4(3m)^2} \biggl\{ \int \bm{dP dq dq' dq_a'' d\xi_a}  \bm{P} \cdot \partialvecright{q_a''}  \biggl[ \Psi'^{*} \Psi   \expcoordmomvec{\xi_a}{q_a''}{q_a'} (\bm{q''_a} + \bm{q'_a}) \cdot \bm{P} \biggr] \\
		& - \int \bm{dP dq dq' dq_a'' d\xi_a} \bm{P} \cdot \partialvecright{q_a'}  \biggl[ \Psi'^{*} \Psi (\bm{q''_a} + \bm{q'_a}) \cdot \bm{P} \biggr] \expcoordmomvec{\xi_a}{q_a''}{q_a'} +  \bigl[ (\bm{q', q''}) \rightarrow (\bm{q'', q})\bigr]  \biggr\}
	\end{split}
\end{equation}}

Since $\Psi$ is a square-integrable function, it is legitimate to assume that the first term in the above equation and its counterpart obtained by substituting
 $(\bm{q', q''}) \rightarrow (\bm{q'', q})$ vanish at the boundary when the integration domain becomes arbitrarily large. By carrying out a partial derivation and by considering all terms in ${\chi_0^{(1)}}$ we arrive again at the result (\ref{boostcorr chizeroone}). 

\section{Orthogonality and closure relations}
    \label{appendix:orthoclosure}
We define physical states in terms of the variables $\bm{p_{\nu}}$, $\bm{r_{\nu}}$, $\bm{\sigma_{\nu}}$, $\bm{\tau_{\nu}}$ as follows. We indicate a generic permutation of indices $1$, $2$, $3$ with $\bm{\alpha_i}=(\alpha_i^{1}, \alpha_i^{2}, \alpha_i^{3})$ and its sign with $\epsilon_{\bm{\alpha_i}}$: 
\begin{equation}
\label{states}
	\begin{split}
		\ket{\bm{p_1}, \sigma_1, \tau_1; \bm{p_2}, \sigma_2, \tau_2; \bm{p_3}, \sigma_3, \tau_3} &=  \sum_{\bm{\alpha_i}} \epsilon_{\bm{\alpha_i}} \ket{\bm{p_{\alpha_i^1}}, \sigma_{\alpha_i^1}, \tau_{\alpha_i^1}; \bm{p_{\alpha_i^2}}, \sigma_{\alpha_i^2}, \tau_{\alpha_i^2};  \bm{p_{\alpha_i^3}}, \sigma_{\alpha_i^3}, \tau_{\alpha_i^3}},\\
		\ket{\bm{r_1}, \sigma_1, \tau_1; \bm{r_2}, \sigma_2, \tau_2; \bm{r_3}, \sigma_3, \tau_3} &=  \sum_{\bm{\alpha_i}} \epsilon_{\bm{\alpha_i}} \ket{\bm{r_{\alpha_i^1}}, \sigma_{\alpha_i^1}, \tau_{\alpha_i^1}; \bm{r_{\alpha_i^2}}, \sigma_{\alpha_i^2}, \tau_{\alpha_i^2};  \bm{r_{\alpha_i^3}}, \sigma_{\alpha_i^3}, \tau_{\alpha_i^3}}.
	\end{split}
\end{equation}
With the shorthand notation $\bm{y_{\nu}}\in\{\bm{r_{\nu}},\bm{p_{\nu}}\}$, $\kappa_{\nu}=(\sigma_{\nu}, \tau_{\nu})$, with $\nu=1, 2, 3$,
orthogonality relations can be written 
\begin{equation}
\label{ortho}
		\braket{\bm{y_1}, \kappa_1; \bm{y_2}, \kappa_2; \bm{y_3}, \kappa_3|\bm{y'_1}, \kappa'_1; \bm{y'_2}, \kappa'_2; \bm{y'_3}, \kappa'_3} = 
		 \prod_{\nu=1}^{3} \delta(\bm{y_{\nu}} - \bm{y'_{\nu}}) \delta_{\kappa_{\nu \nu'}} +  \text{exchange terms} ,
\end{equation}
and closure relations can be written 
\begin{equation}
\label{closure}
	\begin{split}
		\mathbb{1} &= \frac{1}{6}  \biggl[\sum_{\kappa_1 \kappa_2 \kappa_3} \int \bm{dy_1 dy_2 dy_3} \ket{\bm{y_1}, \kappa_1; \bm{y_2}, \kappa_2; \bm{y_3}} \bra{\bm{y_1}, \kappa_1; \bm{y_2}, \kappa_2; \bm{y_3}, \kappa_3}\biggr].
	\end{split}
\end{equation}
In view of the above, Jacobi variables of momentum and position satisfy the following orbital closure and orthogonality relations, where permutation terms are understood: 
\begin{align}
	\label{closurejacobi:P,p}
	\int \bm{dP dq_a dq_b} \ket{\bm{P, q_a, q_b}}\bra{\bm{P, q_a, q_b}} &= \mathbb{1}_{(\bm{P, q_a, q_b})}; \\
	\label{orthojacobi:p}
	\braket{\bm{P, q_a, q_b}| \bm{P', q_a', q_b'}} &= \orthodelta{P}{P'} \orthodelta{q_a}{q_a'} \orthodelta{q_b}{q_b'}; \\
	\label{closurejacobi:R,xi}
	\int \bm{dR d\xi_a d\xi_b} \ket{\bm{R, \xi_a, \xi_b}}\bra{\bm{R, \xi_a, \xi_b}} &= \mathbb{1}_{(\bm{R, \xi_a, \xi_b})}; \\
	\label{orthojacobi:R,xi}
	\braket{\bm{R, \xi_a, \xi_b}| \bm{R', \xi_a', \xi_b'}} &= \orthodelta{R}{R'} \orthodelta{\xi_a}{\xi_a'} \orthodelta{\xi_b}{\xi_b'}.
\end{align}

\end{appendices}
\twocolumn
\bibliography{bib}

\providecommand{\noopsort}[1]{}\providecommand{\singleletter}[1]{#1}

\begin{thebibliography}{36}
\ifx \bisbn   \undefined \def \bisbn  #1{ISBN #1}\fi
\ifx \binits  \undefined \def \binits#1{#1}\fi
\ifx \bauthor  \undefined \def \bauthor#1{#1}\fi
\ifx \batitle  \undefined \def \batitle#1{#1}\fi
\ifx \bjtitle  \undefined \def \bjtitle#1{#1}\fi
\ifx \bvolume  \undefined \def \bvolume#1{\textbf{#1}}\fi
\ifx \byear  \undefined \def \byear#1{#1}\fi
\ifx \bissue  \undefined \def \bissue#1{#1}\fi
\ifx \bfpage  \undefined \def \bfpage#1{#1}\fi
\ifx \blpage  \undefined \def \blpage #1{#1}\fi
\ifx \burl  \undefined \def \burl#1{\textsf{#1}}\fi
\ifx \doiurl  \undefined \def \doiurl#1{\url{https://doi.org/#1}}\fi
\ifx \betal  \undefined \def \betal{\textit{et al.}}\fi
\ifx \binstitute  \undefined \def \binstitute#1{#1}\fi
\ifx \binstitutionaled  \undefined \def \binstitutionaled#1{#1}\fi
\ifx \bctitle  \undefined \def \bctitle#1{#1}\fi
\ifx \beditor  \undefined \def \beditor#1{#1}\fi
\ifx \bpublisher  \undefined \def \bpublisher#1{#1}\fi
\ifx \bbtitle  \undefined \def \bbtitle#1{#1}\fi
\ifx \bedition  \undefined \def \bedition#1{#1}\fi
\ifx \bseriesno  \undefined \def \bseriesno#1{#1}\fi
\ifx \blocation  \undefined \def \blocation#1{#1}\fi
\ifx \bsertitle  \undefined \def \bsertitle#1{#1}\fi
\ifx \bsnm \undefined \def \bsnm#1{#1}\fi
\ifx \bsuffix \undefined \def \bsuffix#1{#1}\fi
\ifx \bparticle \undefined \def \bparticle#1{#1}\fi
\ifx \barticle \undefined \def \barticle#1{#1}\fi
\bibcommenthead
\ifx \bconfdate \undefined \def \bconfdate #1{#1}\fi
\ifx \botherref \undefined \def \botherref #1{#1}\fi
\ifx \url \undefined \def \url#1{\textsf{#1}}\fi
\ifx \bchapter \undefined \def \bchapter#1{#1}\fi
\ifx \bbook \undefined \def \bbook#1{#1}\fi
\ifx \bcomment \undefined \def \bcomment#1{#1}\fi
\ifx \oauthor \undefined \def \oauthor#1{#1}\fi
\ifx \citeauthoryear \undefined \def \citeauthoryear#1{#1}\fi
\ifx \endbibitem  \undefined \def \endbibitem {}\fi
\ifx \bconflocation  \undefined \def \bconflocation#1{#1}\fi
\ifx \arxivurl  \undefined \def \arxivurl#1{\textsf{#1}}\fi
\csname PreBibitemsHook\endcsname

\bibitem[\protect\citeauthoryear{Weinberg}{1990}]{Weinberg:1990rz}
\begin{barticle}
\bauthor{\bsnm{Weinberg}, \binits{S.}}:
\batitle{{Nuclear forces from chiral Lagrangians}}.
\bjtitle{Phys. Lett. B}
\bvolume{251},
\bfpage{288}--\blpage{292}
(\byear{1990})
\doiurl{10.1016/0370-2693(90)90938-3}
\end{barticle}
\endbibitem

\bibitem[\protect\citeauthoryear{Weinberg}{1991}]{Weinberg:1991um}
\begin{barticle}
\bauthor{\bsnm{Weinberg}, \binits{S.}}:
\batitle{{Effective chiral Lagrangians for nucleon - pion interactions and
  nuclear forces}}.
\bjtitle{Nucl. Phys. B}
\bvolume{363},
\bfpage{3}--\blpage{18}
(\byear{1991})
\doiurl{10.1016/0550-3213(91)90231-L}
\end{barticle}
\endbibitem

\bibitem[\protect\citeauthoryear{Weinberg}{1992}]{Weinberg:1992yk}
\begin{barticle}
\bauthor{\bsnm{Weinberg}, \binits{S.}}:
\batitle{{Three body interactions among nucleons and pions}}.
\bjtitle{Phys. Lett. B}
\bvolume{295},
\bfpage{114}--\blpage{121}
(\byear{1992})
\doiurl{10.1016/0370-2693(92)90099-P}
{\href{https://arxiv.org/abs/hep-ph/9209257}{{arXiv:hep-ph/9209257}}}
\end{barticle}
\endbibitem

\bibitem[\protect\citeauthoryear{Bedaque and van Kolck}{2002}]{Bedaque:2002mn}
\begin{barticle}
\bauthor{\bsnm{Bedaque}, \binits{P.F.}},
\bauthor{\bsnm{Kolck}, \binits{U.}}:
\batitle{{Effective field theory for few nucleon systems}}.
\bjtitle{Ann. Rev. Nucl. Part. Sci.}
\bvolume{52},
\bfpage{339}--\blpage{396}
(\byear{2002})
\doiurl{10.1146/annurev.nucl.52.050102.090637}
{\href{https://arxiv.org/abs/nucl-th/0203055}{{arXiv:nucl-th/0203055}}}
\end{barticle}
\endbibitem

\bibitem[\protect\citeauthoryear{Epelbaum}{2006}]{Epelbaum:2005pn}
\begin{barticle}
\bauthor{\bsnm{Epelbaum}, \binits{E.}}:
\batitle{{Few-nucleon forces and systems in chiral effective field theory}}.
\bjtitle{Prog. Part. Nucl. Phys.}
\bvolume{57},
\bfpage{654}--\blpage{741}
(\byear{2006})
\doiurl{10.1016/j.ppnp.2005.09.002}
{\href{https://arxiv.org/abs/nucl-th/0509032}{{arXiv:nucl-th/0509032}}}
\end{barticle}
\endbibitem

\bibitem[\protect\citeauthoryear{Epelbaum et~al.}{2009}]{Epelbaum:2008ga}
\begin{barticle}
\bauthor{\bsnm{Epelbaum}, \binits{E.}},
\bauthor{\bsnm{Hammer}, \binits{H.-W.}},
\bauthor{\bsnm{Meissner}, \binits{U.-G.}}:
\batitle{{Modern Theory of Nuclear Forces}}.
\bjtitle{Rev. Mod. Phys.}
\bvolume{81},
\bfpage{1773}--\blpage{1825}
(\byear{2009})
\doiurl{10.1103/RevModPhys.81.1773}
{\href{https://arxiv.org/abs/0811.1338}{{arXiv:0811.1338}}}
{[nucl-th]}
\end{barticle}
\endbibitem

\bibitem[\protect\citeauthoryear{Machleidt and Entem}{2011}]{Machleidt:2011zz}
\begin{barticle}
\bauthor{\bsnm{Machleidt}, \binits{R.}},
\bauthor{\bsnm{Entem}, \binits{D.R.}}:
\batitle{{Chiral effective field theory and nuclear forces}}.
\bjtitle{Phys. Rept.}
\bvolume{503},
\bfpage{1}--\blpage{75}
(\byear{2011})
\doiurl{10.1016/j.physrep.2011.02.001}
{\href{https://arxiv.org/abs/1105.2919}{{arXiv:1105.2919}}}
{[nucl-th]}
\end{barticle}
\endbibitem

\bibitem[\protect\citeauthoryear{Jenkins and Manohar}{1991}]{Jenkins:1990jv}
\begin{barticle}
\bauthor{\bsnm{Jenkins}, \binits{E.E.}},
\bauthor{\bsnm{Manohar}, \binits{A.V.}}:
\batitle{{Baryon chiral perturbation theory using a heavy fermion Lagrangian}}.
\bjtitle{Phys. Lett. B}
\bvolume{255},
\bfpage{558}--\blpage{562}
(\byear{1991})
\doiurl{10.1016/0370-2693(91)90266-S}
\end{barticle}
\endbibitem

\bibitem[\protect\citeauthoryear{Bernard et~al.}{1992}]{Bernard:1992qa}
\begin{barticle}
\bauthor{\bsnm{Bernard}, \binits{V.}},
\bauthor{\bsnm{Kaiser}, \binits{N.}},
\bauthor{\bsnm{Kambor}, \binits{J.}},
\bauthor{\bsnm{Meissner}, \binits{U.G.}}:
\batitle{{Chiral structure of the nucleon}}.
\bjtitle{Nucl. Phys. B}
\bvolume{388},
\bfpage{315}--\blpage{345}
(\byear{1992})
\doiurl{10.1016/0550-3213(92)90615-I}
\end{barticle}
\endbibitem

\bibitem[\protect\citeauthoryear{Foldy}{1961}]{Foldy:1960nb}
\begin{barticle}
\bauthor{\bsnm{Foldy}, \binits{L.L.}}:
\batitle{{Relativistic particle systems with interactions}}.
\bjtitle{Phys. Rev.}
\bvolume{122},
\bfpage{275}--\blpage{288}
(\byear{1961})
\doiurl{10.1103/PhysRev.122.275}
\end{barticle}
\endbibitem

\bibitem[\protect\citeauthoryear{Krajcik and Foldy}{1974}]{Krajcik:1974nv}
\begin{barticle}
\bauthor{\bsnm{Krajcik}, \binits{R.A.}},
\bauthor{\bsnm{Foldy}, \binits{L.L.}}:
\batitle{{Relativistic center-of-mass variables for composite systems with
  arbitrary internal interactions}}.
\bjtitle{Phys. Rev. D}
\bvolume{10},
\bfpage{1777}--\blpage{1795}
(\byear{1974})
\doiurl{10.1103/PhysRevD.10.1777}
\end{barticle}
\endbibitem

\bibitem[\protect\citeauthoryear{Girlanda et~al.}{2010}]{GirlandaSchiavilla}
\begin{barticle}
\bauthor{\bsnm{Girlanda}, \binits{L.}},
\bauthor{\bsnm{Pastore}, \binits{S.}},
\bauthor{\bsnm{Schiavilla}, \binits{R.}},
\bauthor{\bsnm{Viviani}, \binits{M.}}:
\batitle{Relativity constraints on the two-nucleon contact interaction}.
\bjtitle{Phys. Rev. C}
\bvolume{81},
\bfpage{034005}
(\byear{2010})
\doiurl{10.1103/PhysRevC.81.034005}
\end{barticle}
\endbibitem

\bibitem[\protect\citeauthoryear{Girlanda and Viviani}{2011}]{Girlanda:2010zz}
\begin{barticle}
\bauthor{\bsnm{Girlanda}, \binits{L.}},
\bauthor{\bsnm{Viviani}, \binits{M.}}:
\batitle{{Relativistic Covariance of the 2-Nucleon Contact Interactions}}.
\bjtitle{Few Body Syst.}
\bvolume{49},
\bfpage{51}--\blpage{60}
(\byear{2011})
\doiurl{10.1007/s00601-010-0185-6}
\end{barticle}
\endbibitem

\bibitem[\protect\citeauthoryear{Girlanda et~al.}{2020}]{unitarity}
\begin{barticle}
\bauthor{\bsnm{Girlanda}, \binits{L.}},
\bauthor{\bsnm{Kievsky}, \binits{A.}},
\bauthor{\bsnm{Marcucci}, \binits{L.E.}},
\bauthor{\bsnm{Viviani}, \binits{M.}}:
\batitle{{Unitary ambiguity of NN contact interactions and the 3N force}}.
\bjtitle{Phys. Rev. C}
\bvolume{102},
\bfpage{064003}
(\byear{2020})
\doiurl{10.1103/PhysRevC.102.064003}
{\href{https://arxiv.org/abs/2007.04161}{{arXiv:2007.04161}}}
{[nucl-th]}
\end{barticle}
\endbibitem

\bibitem[\protect\citeauthoryear{Filandri and
  Girlanda}{2023}]{Filandri:2023qio}
\begin{barticle}
\bauthor{\bsnm{Filandri}, \binits{E.}},
\bauthor{\bsnm{Girlanda}, \binits{L.}}:
\batitle{{Momentum dependent nucleon-nucleon contact interaction from a
  relativistic Lagrangian}}.
\bjtitle{Phys. Lett. B}
\bvolume{841},
\bfpage{137957}
(\byear{2023})
\doiurl{10.1016/j.physletb.2023.137957}
{\href{https://arxiv.org/abs/2303.10084}{{arXiv:2303.10084}}}
{[nucl-th]}
\end{barticle}
\endbibitem

\bibitem[\protect\citeauthoryear{Girlanda et~al.}{2011}]{Girlanda:2011fh}
\begin{barticle}
\bauthor{\bsnm{Girlanda}, \binits{L.}},
\bauthor{\bsnm{Kievsky}, \binits{A.}},
\bauthor{\bsnm{Viviani}, \binits{M.}}:
\batitle{{Subleading contributions to the three-nucleon contact interaction}}.
\bjtitle{Phys. Rev. C}
\bvolume{84}(\bissue{1}),
\bfpage{014001}
(\byear{2011})
\doiurl{10.1103/PhysRevC.84.014001}
{\href{https://arxiv.org/abs/1102.4799}{{arXiv:1102.4799}}}
{[nucl-th]}.
\bcomment{[Erratum: Phys.Rev.C 102, 019903 (2020)]}
\end{barticle}
\endbibitem

\bibitem[\protect\citeauthoryear{Girlanda et~al.}{2019}]{Girlanda:2018xrw}
\begin{barticle}
\bauthor{\bsnm{Girlanda}, \binits{L.}},
\bauthor{\bsnm{Kievsky}, \binits{A.}},
\bauthor{\bsnm{Viviani}, \binits{M.}},
\bauthor{\bsnm{Marcucci}, \binits{L.E.}}:
\batitle{{Short-range three-nucleon interaction from A=3 data and its
  hierarchical structure}}.
\bjtitle{Phys. Rev. C}
\bvolume{99}(\bissue{5}),
\bfpage{054003}
(\byear{2019})
\doiurl{10.1103/PhysRevC.99.054003}
{\href{https://arxiv.org/abs/1811.09398}{{arXiv:1811.09398}}}
{[nucl-th]}
\end{barticle}
\endbibitem

\bibitem[\protect\citeauthoryear{Wita\l{}a et~al.}{2022}]{Witala:2022rzl}
\begin{barticle}
\bauthor{\bsnm{Wita\l{}a}, \binits{H.}},
\bauthor{\bsnm{Golak}, \binits{J.}},
\bauthor{\bsnm{Skibi\'nski}, \binits{R.}}:
\batitle{{Significance of chiral three-nucleon force contact terms for
  understanding of elastic nucleon-deuteron scattering}}.
\bjtitle{Phys. Rev. C}
\bvolume{105}(\bissue{5}),
\bfpage{054004}
(\byear{2022})
\doiurl{10.1103/PhysRevC.105.054004}
{\href{https://arxiv.org/abs/2203.08499}{{arXiv:2203.08499}}}
{[nucl-th]}
\end{barticle}
\endbibitem

\bibitem[\protect\citeauthoryear{Reinert et~al.}{2017}]{Reinert}
\begin{barticle}
\bauthor{\bsnm{Reinert}, \binits{P.}},
\bauthor{\bsnm{Krebs}, \binits{H.}},
\bauthor{\bsnm{Epelbaum}, \binits{E.}}:
\batitle{Semilocal momentum-space regularized chiral two-nucleon potentials up
  to fifth order}.
\bjtitle{Eur. Phys. J. A}
\bvolume{54},
\bfpage{1}--\blpage{49}
(\byear{2017})
\end{barticle}
\endbibitem

\bibitem[\protect\citeauthoryear{Girlanda et~al.}{2023}]{Girlanda:2023znc}
\begin{botherref}
\oauthor{\bsnm{Girlanda}, \binits{L.}},
\oauthor{\bsnm{Filandri}, \binits{E.}},
\oauthor{\bsnm{Kievsky}, \binits{A.}},
\oauthor{\bsnm{Marcucci}, \binits{L.E.}},
\oauthor{\bsnm{Viviani}, \binits{M.}}:
{Effect of the N3LO three-nucleon contact interaction on p-d scattering
  observables}
(2023)
{\href{https://arxiv.org/abs/2302.03468}{{arXiv:2302.03468}}}
{[nucl-th]}
\end{botherref}
\endbibitem

\bibitem[\protect\citeauthoryear{Fisher et~al.}{2006}]{Fisher_2006}
\begin{botherref}
\oauthor{\bsnm{Fisher}, \binits{B.M.}},
\oauthor{\bsnm{Brune}, \binits{C.R.}},
\oauthor{\bsnm{Karwowski}, \binits{H.J.}},
\oauthor{\bsnm{Leonard}, \binits{D.S.}},
\oauthor{\bsnm{Ludwig}, \binits{E.J.}},
\oauthor{\bsnm{Black}, \binits{T.C.}},
\oauthor{\bsnm{Viviani}, \binits{M.}},
\oauthor{\bsnm{Kievsky}, \binits{A.}},
\oauthor{\bsnm{Rosati}, \binits{S.}}:
Proton-3he elastic scattering at low energies.
Phys. Rev. C
\textbf{74}(3)
(2006)
\doiurl{10.1103/physrevc.74.034001}
\end{botherref}
\endbibitem

\bibitem[\protect\citeauthoryear{Deltuva and
  Fonseca}{2007a}]{PhysRevC.75.014005}
\begin{barticle}
\bauthor{\bsnm{Deltuva}, \binits{A.}},
\bauthor{\bsnm{Fonseca}, \binits{A.C.}}:
\batitle{Four-nucleon scattering: Ab initio calculations in momentum space}.
\bjtitle{Phys. Rev. C}
\bvolume{75},
\bfpage{014005}
(\byear{2007})
\doiurl{10.1103/PhysRevC.75.014005}
\end{barticle}
\endbibitem

\bibitem[\protect\citeauthoryear{Deltuva and Fonseca}{2007b}]{Deltuva_2007}
\begin{botherref}
\oauthor{\bsnm{Deltuva}, \binits{A.}},
\oauthor{\bsnm{Fonseca}, \binits{A.C.}}:
Four-body calculation of proton-3he scattering.
Phys. Rev. Lett.
\textbf{98}(16)
(2007)
\doiurl{10.1103/physrevlett.98.162502}
\end{botherref}
\endbibitem

\bibitem[\protect\citeauthoryear{Deltuva and
  Fonseca}{2007c}]{PhysRevC.76.021001}
\begin{barticle}
\bauthor{\bsnm{Deltuva}, \binits{A.}},
\bauthor{\bsnm{Fonseca}, \binits{A.C.}}:
\batitle{Ab initio four-body calculation of $n$-$^{3}\mathrm{He}$,
  $p$-$^{3}\mathrm{H}$, and $d$-$d$ scattering}.
\bjtitle{Phys. Rev. C}
\bvolume{76},
\bfpage{021001}
(\byear{2007})
\doiurl{10.1103/PhysRevC.76.021001}
\end{barticle}
\endbibitem

\bibitem[\protect\citeauthoryear{Fettes et~al.}{1998}]{Fettes:1998ud}
\begin{barticle}
\bauthor{\bsnm{Fettes}, \binits{N.}},
\bauthor{\bsnm{Meissner}, \binits{U.-G.}},
\bauthor{\bsnm{Steininger}, \binits{S.}}:
\batitle{{Pion - nucleon scattering in chiral perturbation theory. 1. Isospin
  symmetric case}}.
\bjtitle{Nucl. Phys. A}
\bvolume{640},
\bfpage{199}--\blpage{234}
(\byear{1998})
\doiurl{10.1016/S0375-9474(98)00452-7}
{\href{https://arxiv.org/abs/hep-ph/9803266}{{arXiv:hep-ph/9803266}}}
\end{barticle}
\endbibitem

\bibitem[\protect\citeauthoryear{Fettes et~al.}{2000}]{Fettes:2000gb}
\begin{barticle}
\bauthor{\bsnm{Fettes}, \binits{N.}},
\bauthor{\bsnm{Meissner}, \binits{U.-G.}},
\bauthor{\bsnm{Mojzis}, \binits{M.}},
\bauthor{\bsnm{Steininger}, \binits{S.}}:
\batitle{{The Chiral effective pion nucleon Lagrangian of order p**4}}.
\bjtitle{Annals Phys.}
\bvolume{283},
\bfpage{273}--\blpage{302}
(\byear{2000})
\doiurl{10.1006/aphy.2000.6059}
{\href{https://arxiv.org/abs/hep-ph/0001308}{{arXiv:hep-ph/0001308}}}.
\bcomment{[Erratum: Annals Phys. 288, 249--250 (2001)]}
\end{barticle}
\endbibitem

\bibitem[\protect\citeauthoryear{Xiao et~al.}{2019}]{Xiao:2018jot}
\begin{barticle}
\bauthor{\bsnm{Xiao}, \binits{Y.}},
\bauthor{\bsnm{Geng}, \binits{L.-S.}},
\bauthor{\bsnm{Ren}, \binits{X.-L.}}:
\batitle{{Covariant chiral nucleon-nucleon contact Lagrangian up to order
  $\mathcal{O}(q^4)$}}.
\bjtitle{Phys. Rev. C}
\bvolume{99}(\bissue{2}),
\bfpage{024004}
(\byear{2019})
\doiurl{10.1103/PhysRevC.99.024004}
{\href{https://arxiv.org/abs/1812.03005}{{arXiv:1812.03005}}}
{[nucl-th]}
\end{barticle}
\endbibitem

\bibitem[\protect\citeauthoryear{Petschauer and
  Kaiser}{2013}]{Petschauer:2013uua}
\begin{barticle}
\bauthor{\bsnm{Petschauer}, \binits{S.}},
\bauthor{\bsnm{Kaiser}, \binits{N.}}:
\batitle{{Relativistic SU(3) chiral baryon-baryon Lagrangian up to order
  $q^{2}$}}.
\bjtitle{Nucl. Phys. A}
\bvolume{916},
\bfpage{1}--\blpage{29}
(\byear{2013})
\doiurl{10.1016/j.nuclphysa.2013.07.010}
{\href{https://arxiv.org/abs/1305.3427}{{arXiv:1305.3427}}}
{[nucl-th]}
\end{barticle}
\endbibitem

\bibitem[\protect\citeauthoryear{Georgi}{1991}]{Georgi:1991ch}
\begin{barticle}
\bauthor{\bsnm{Georgi}, \binits{H.}}:
\batitle{{On-shell effective field theory}}.
\bjtitle{Nucl. Phys. B}
\bvolume{361},
\bfpage{339}--\blpage{350}
(\byear{1991})
\doiurl{10.1016/0550-3213(91)90244-R}
\end{barticle}
\endbibitem

\bibitem[\protect\citeauthoryear{Arzt}{1995}]{Arzt:1993gz}
\begin{barticle}
\bauthor{\bsnm{Arzt}, \binits{C.}}:
\batitle{{Reduced effective Lagrangians}}.
\bjtitle{Phys. Lett. B}
\bvolume{342},
\bfpage{189}--\blpage{195}
(\byear{1995})
\doiurl{10.1016/0370-2693(94)01419-D}
{\href{https://arxiv.org/abs/hep-ph/9304230}{{arXiv:hep-ph/9304230}}}
\end{barticle}
\endbibitem

\bibitem[\protect\citeauthoryear{Forest et~al.}{1994}]{Forest:1994mw}
\begin{botherref}
\oauthor{\bsnm{Forest}, \binits{J.L.}},
\oauthor{\bsnm{Pandharipande}, \binits{V.R.}},
\oauthor{\bsnm{Friar}, \binits{J.L.}}:
{Pedagogical studies of relativistic Hamiltonians}
(1994)
{\href{https://arxiv.org/abs/nucl-th/9410011}{{arXiv:nucl-th/9410011}}}
\end{botherref}
\endbibitem

\bibitem[\protect\citeauthoryear{Dirac}{1949}]{Dirac:1949cp}
\begin{barticle}
\bauthor{\bsnm{Dirac}, \binits{P.A.M.}}:
\batitle{{Forms of Relativistic Dynamics}}.
\bjtitle{Rev. Mod. Phys.}
\bvolume{21},
\bfpage{392}--\blpage{399}
(\byear{1949})
\doiurl{10.1103/RevModPhys.21.392}
\end{barticle}
\endbibitem

\bibitem[\protect\citeauthoryear{Friar}{1975}]{PhysRevC.12.695}
\begin{barticle}
\bauthor{\bsnm{Friar}, \binits{J.L.}}:
\batitle{Relativistic effects on the wave function of a moving system}.
\bjtitle{Phys. Rev. C}
\bvolume{12},
\bfpage{695}--\blpage{698}
(\byear{1975})
\doiurl{10.1103/PhysRevC.12.695}
\end{barticle}
\endbibitem

\bibitem[\protect\citeauthoryear{Carlson et~al.}{1993}]{Carlson:1993zz}
\begin{barticle}
\bauthor{\bsnm{Carlson}, \binits{J.}},
\bauthor{\bsnm{Pandharipande}, \binits{V.R.}},
\bauthor{\bsnm{Schiavilla}, \binits{R.}}:
\batitle{{Variational Monte Carlo calculations of H-3 and He-4 with a
  relativistic Hamiltonian}}.
\bjtitle{Phys. Rev. C}
\bvolume{47},
\bfpage{484}--\blpage{497}
(\byear{1993})
\doiurl{10.1103/PhysRevC.47.484}
\end{barticle}
\endbibitem

\bibitem[\protect\citeauthoryear{Forest et~al.}{1995}]{Forest:1995zz}
\begin{barticle}
\bauthor{\bsnm{Forest}, \binits{J.L.}},
\bauthor{\bsnm{Pandharipande}, \binits{V.R.}},
\bauthor{\bsnm{Carlson}, \binits{J.}},
\bauthor{\bsnm{Schiavilla}, \binits{R.}}:
\batitle{{Variational Monte Carlo calculations of H-3 and He-4 with a
  relativistic Hamiltonian}}.
\bjtitle{Phys. Rev. C}
\bvolume{52},
\bfpage{576}--\blpage{577}
(\byear{1995})
\doiurl{10.1103/PhysRevC.52.576}
\end{barticle}
\endbibitem

\bibitem[\protect\citeauthoryear{Nasoni}{2022}]{nasoni}
\begin{botherref}
\oauthor{\bsnm{Nasoni}, \binits{A.}}
Laurea Thesis, Universit\`a del Salento, Lecce, Italy
(2022)
\end{botherref}
\endbibitem

\end{thebibliography}
\end{document}